\documentclass[aps,groupaddress,preprint,showpacs,letterpaper,tighten,float,longbibliography,nofootinbib]{revtex4-1}
\usepackage{amssymb}
\usepackage{amsbsy}
\usepackage{amsmath}
\usepackage{epsfig}
\usepackage{wrapfig}
\usepackage{graphicx}
\usepackage{array}
\usepackage{textcomp}
\usepackage{color}
\usepackage{braket}
\usepackage{bm,hyperref}
\usepackage[titletoc,toc,title]{appendix}
\usepackage[normalem]{ulem}
\newcommand{\be}{\begin{equation} }
\newcommand{\ee}{\end{equation} }
\newcommand{\ba}{\begin{eqnarray} }
\newcommand{\ea}{\end{eqnarray} }
\newcommand{\n}{\nonumber \\ }

\newcommand{\mac}{\mathcal}

\definecolor{myblue}{rgb}{.93, .93, 1}

\newcommand{\beq}{\begin{equation}}
\newcommand{\eeq}{\end{equation}}

\newcommand{\figref}[1]{Fig.\,\ref{#1}}
\newcommand{\eqnref}[1]{Eq.\,(\ref{#1})}
\newcommand{\secref}[1]{Section \,\ref{#1}}
\newcommand{\sfigref}[2]{Fig.\,\hyperref[#1]{\ref{#1}(#2)}}
\newcommand{\Ref}[1]{Ref.\,\onlinecite{#1}}

\begin{document}

\title{Fracton Phases of Matter}

\author{Michael Pretko}
\affiliation{Department of Physics and Center for Theory of Quantum Matter, University of Colorado, Boulder, CO 80309}

\author{Xie Chen}
\affiliation{Department of Physics and Institute for Quantum Information and Matter, California Institute of Technology, Pasadena, CA 91125}

\author{Yizhi You}
\affiliation{Princeton Center for Theoretical Science, Princeton University, NJ 08544, USA}

\date{\today}

\begin{abstract}
Fractons are a new type of quasiparticle which are immobile in isolation, but can often move by forming bound states.  Fractons are found in a variety of physical settings, such as spin liquids and elasticity theory, and exhibit unusual phenomenology, such as gravitational physics and localization.  The past several years have seen a surge of interest in these exotic particles, which have come to the forefront of modern condensed matter theory.  In this review, we provide a broad treatment of fractons, ranging from pedagogical introductory material to discussions of recent advances in the field.  We begin by demonstrating how the fracton phenomenon naturally arises as a consequence of higher moment conservation laws, often accompanied by the emergence of tensor gauge theories.  We then provide a survey of fracton phases in spin models, along with the various tools used to characterize them, such as the foliation framework.  We discuss in detail the manifestation of fracton physics in elasticity theory, as well as the connections of fractons with localization and gravitation.  Finally, we provide an overview of some recently proposed platforms for fracton physics, such as Majorana islands and hole-doped antiferromagnets.  We conclude with some open questions and an outlook on the field.
\end{abstract}

\maketitle

\tableofcontents

\section{Introduction}
\label{intro}=

The field of condensed matter physics studies the complex and often surprising collective behavior of systems containing many particles.  One of the most striking examples of new physics which arises in such many-body systems is the concept of an emergent quasiparticle.  Strong interactions between the microscopic particles can often drive the formation of emergent quasiparticle excitations with vastly different properties from any known fundamental particle.  The concept of a quasiparticle dates back to Landau's theory of Fermi liquids, in which interactions between electrons lead to the formation of quasiparticle excitations with the same charge as an electron, but with a different mass.  A more dramatic example of an emergent quasiparticle was later found in the context of fractional quantum Hall systems, where Laughlin quasiparticles carry only a fraction of the elementary electric charge.  Since then, a wide array of quasiparticles has been discovered, often possessing fractionalized quantum numbers or anyonic quantum statistics.

\begin{figure}
    \centering
    \includegraphics[width=1\columnwidth]{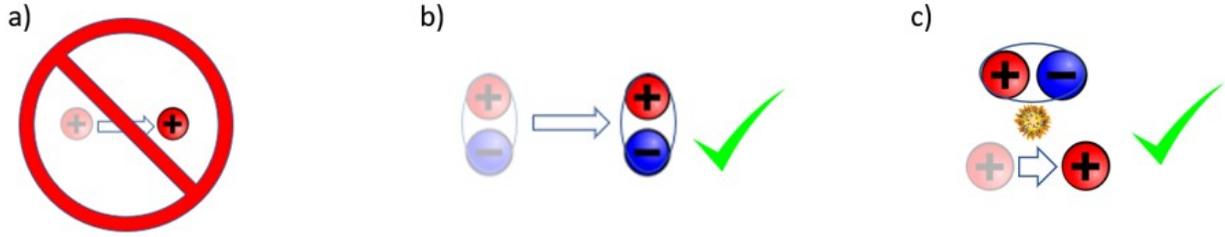}
    \caption{a) A single fracton cannot move freely in any direction.  b) Fractons can sometimes move by forming certain bound states, such as dipoles.  c) It is also possible for a fracton to move at the expense of creating new particles out of the vacuum.}
    \label{fig:fracton}
\end{figure}

Recently, however, a new type of emergent quasiparticle has been encountered which differs from all previously known particles in an unusual way.  Fractons are quasiparticles which lack an ability previously assumed to be inherent to all particles: namely the ability to move.  A fracton is a quasiparticle which, in isolation, is unable to move in response to an applied force \cite{chamon,haah,vhf1,vhf2,sub}.  However, depending on the details of the model, fractons can sometimes move by combining to form certain bound states, as depicted in Figure \ref{fig:fracton}.  Fracton models are often classified as ``type-I" if they possess stable mobile bound states, and as ``type-II" if all mobile bound states can decay directly into the vacuum \cite{vhf2}.  It is also possible for an individual fracton to move at the cost of creating new fractons out of the vacuum at each step of its motion.  However, in the absence of a constant energy input to sustain this particle creation, an individual fracton will remain immobile.  These unusual new particles were first encountered in certain exactly-solvable three-dimensional spin and Majorana models \cite{chamon,castelnovo,haah,vhf1,vhf2,bravyi2,yoshida}, but have since been shown to arise in contexts ranging from topological crystalline defects \cite{elasticity} to plaquette-ordered paramagnets \cite{plaquette} (see also precursor work in Ref. \cite{rsvp}).  Furthermore, the restricted mobility of fractons causes them to exhibit a variety of unusual properties, such as nonergodic behavior \cite{glassy,localization} and even gravitational physics \cite{mach,holo1}.  At a practical level, there is hope that the immobility of fractons may even be harnessed for the purposes of quantum information storage \cite{haah,bravyi,terhal,parallel}.

The field of fractons has a somewhat complicated history, and we give only a brief overview.  It is generally agreed upon that the first manifestation of fracton behavior was encountered in a spin model exhibiting glassy dynamics constructed by Chamon \cite{chamon}, though there is also important conceptual overlap between fractons and earlier work on kinetically constrained models \cite{ritort,newmanmoore,garrahan}.  Later, Haah designed the paradigmatic type-II fracton model, featuring a characteristic fractal structure, with the goal of creating a self-correcting quantum memory \cite{haah}.  However, the significance of these two models, often known as the Chamon model and Haah's code respectively, was not immediately appreciated.  It was not until the seminal work of Vijay, Haah, and Fu that it became clear that these models were only two examples of a much larger class of fracton systems, representing a fundamentally new type of phase of matter \cite{vhf1,vhf2}.  Vijay, Haah, and Fu constructed several now-prototypical fracton models in three dimensions, such as the X-cube model.  Additionally, they recognized the existence of several close cousins of fractons: particles which can only move along a one- or two-dimensional subspace of a three-dimensional system.  These particles have since come to be known as lineons and planons respectively, or sometimes more generally as subdimensional particles.

The next major advance in the understanding of fractons came with the realization by one of the present authors (MP) that the restricted mobility of fractons can be naturally understood in terms of a set of higher moment conservation laws, which often arise as a consequence of an emergent symmetric tensor gauge theory \cite{sub,genem}.  For example, the simplest such gauge theories feature conservation of both charge and dipole moment, which immobilizes individual charges but allows for motion of stable dipolar bound states.  Building on earlier work on symmetric tensor gauge theories \cite{xu,rasmussen,wen1,wen2,horava}, MP showed that these gauge theories provide an effective description of a broad class of fracton phases featuring emergent gapless gauge modes.  It was later shown by Ma, Hermele, and Chen \cite{higgs1}, and independently by Bulmash and Barkeshli \cite{higgs2}, that certain symmetric tensor gauge theories give rise to the previously studied gapped fracton models via the Higgs mechanism.  From this viewpoint, various spin-1/2 fracton models can be understood as types of $Z_2$ symmetric tensor gauge theories.  In addition to shedding internal light on the field of fractons, the symmetric tensor gauge theory formalism has also drawn unexpected connections between fractons and other areas of physics, such as elasticity theory \cite{elasticity} and gravity \cite{mach}.  Due to their key role in the field of fractons, we begin by discussing some basic aspects of symmetric tensor gauge theories in Section \ref{sec:tensor}.

Recently, there has been further significant progress on the understanding of fracton phases in gapped spin models.  Useful tools have now been developed for relating such fracton phases to more familiar topological phases of matter.  For example, it has been shown how certain three-dimensional fracton phases can arise via strongly coupling together layers of two-dimensional topological phases \cite{hanlayer,sagarlayer}.  Various schemes have also been proposed for generalizing the string-net condensate picture for ordinary topological phases to fracton phases \cite{cage,smn}.  In Section \ref{sec:spin}, we describe various aspects of fractons in spin models, beginning with a description of some prototypical models, such as the X-cube model and Haah's code, and ending with a discussion of more recent developments.  In Section \ref{sec:foliation}, we provide a separate discussion of the important topic of the classification of gapped fracton phases, with special emphasis on the well-developed foliation framework, as pioneered by Shirley, Slagle, and Chen \cite{fol1,fol2,fol3,fol4,fol5,fol6}.  We also describe some other recent tools developed for characterizing fracton phases, such as the Pai-Hermele theory of fusion and braiding in fracton systems \cite{fusion}.

While much of the work on fractons takes place in the context of abstract spin models and gauge theories, it is important to note that fracton physics has a very concrete realization as the topological lattice defects of ordinary crystals.  Specifically, the disclinations and dislocations of two-dimensional crystals exhibit the restricted mobility of fractons and lineons, respectively.  This connection is made precise via a duality transformation, often referred to as ``fracton-elasticity duality," which maps the elasticity theory of crystals onto a symmetric tensor gauge theory \cite{elasticity}.  We discuss this duality in detail in Section \ref{sec:elasticity}, along with its various generalizations \cite{3delasticity,gromov,potter,z3,supersolid,elasticityprb,cosserat}.  For example, the duality can be extended to three-dimensional elasticity theory, giving rise to the concept of fractonic lines, i.e line-like excitations without the ability to move \cite{3delasticity}.

In the following sections, we discuss some of the phenomenology of fractons, which is important for the detection of fracton behavior in experiments.  As a first notable example, the immobility of fractons serves as a significant impediment to thermalization.  Fracton physics generically causes systems to be slow to reach thermal equilibrium, in a manifestation of glassy dynamics, as studied first by Chamon \cite{chamon} and more systematically by Prem et al. \cite{glassy}.  In certain cases, fracton systems can exhibit truly non-ergodic behavior, failing to ever reach thermal equilibrium, as shown by Pai et al. \cite{localization}.  In Section \ref{sec:nonergodic}, we describe these unusual thermodynamic aspects of fracton systems.  Another unusual characteristic of fracton systems is that, depending on the precise form of the conservation laws, fractons can often exhibit gravitational behavior, in the sense of a universal attractive force between particles which is encoded in an effective geometry \cite{mach}.  We describe how this gravitational physics arises out of the tensor gauge theory formalism in Section \ref{sec:gravity}.

In Section \ref{sec:cmp}, we move on to discuss various other physical realizations of fractons.  This includes both artificially engineered fractons, for example built from Majorana islands \cite{majorana}, and realizations of fractons in familiar condensed matter settings, such as plaquette paramagnets \cite{plaquette} and hole-doped antiferromagnets \cite{polaron}.  Finally, in Section \ref{sec:conc}, we conclude with some open questions and an outlook on the field of fractons.

\section{Tensor Gauge Theories and Higher Moment Conservation Laws}
\label{sec:tensor}

\subsection{Basic Principles}

While the restricted mobility of fractons may seem unusual at first glance, the basic principles governing their phenomenology can be understood in terms of a remarkably simple class of theories.  Specifically, fracton behavior is seen to arise in gauge theories featuring a symmetric tensor gauge variable.  To illustrate the main idea, it is useful to focus on the simplest type of symmetric tensor gauge theory in three dimensions, corresponding to a generalized Maxwell theory in which the familiar vector potential $\Vec{A}$ is replaced by a rank-2 symmetric tensor potential $A_{ij}$ (where Roman indices refer to spatial coordinates) \cite{xu,rasmussen,sub,genem}.  While this initial discussion may seem abstract, we will discuss in detail later how these effective theories can arise from microscopic models, such as spin Hamiltonians.

To construct a symmetric tensor version of Maxwell theory, it is useful to first specify the gauge transformation, which will largely dictate the form of the rest of the theory.  For Maxwell theory, the pure gauge sector is invariant under transformations of the form $\Vec{A}\rightarrow\Vec{A} + \Vec{\nabla}\alpha$.  (We will return later to the role of the scalar potential $\phi$, which becomes important within the charge sector.)  A natural (though not unique) choice for the tensor Maxwell theory is to specify that the gauge sector must be invariant under:
\begin{equation}
A_{ij}\rightarrow A_{ij} + \partial_i\partial_j\alpha
\end{equation}
For reasons which will become clear later, this theory is typically known as the ``scalar charge theory" \cite{sub}.  Given this gauge transformation, we now wish to construct gauge-invariant field operators, playing the role of electric and magnetic fields.  To obtain an electric field, it is simplest to work in the Hamiltonian formalism.  In this case, we can simply \emph{define} a symmetric tensor electric field $E_{ij}$ as the canonical conjugate to $A_{ij}$, in analogy with the conjugate relationship between $\Vec{A}$ and $\Vec{E}$ in Maxwell theory.  More precisely, we write:
\begin{equation}
[A_{ij}(x),E_{k\ell}(y)] = i(\delta_{ik}\delta_{j\ell} + \delta_{i\ell}\delta_{jk})\delta(x-y)
\end{equation}
where the right-hand side has taken into account the symmetric property of the tensors.

To construct a magnetic field operator, we can simply take a curl on either of the indices of $A_{ij}$ as follows:
\begin{equation}
B_{ij} = \epsilon_{ik\ell}\partial^k A^{\ell}_{\,\,\,j}
\end{equation}
where we are summing over repeated indices.  (Furthermore, we work exclusively in flat space, raising and lowering indices via the flat metric $\delta^{ij}$.)  Note that, in contrast to $A_{ij}$ and $E_{ij}$, the magnetic field tensor $B_{ij}$ is neither symmetric nor antisymmetric in its indices, but does obey the tracelessness condition $B^i_{\,\,i} = 0$.\footnote{While one could symmetrize the magnetic tensor as $\Tilde{B}_{ij} = B_{ij}+B_{ji}$, doing so would fine-tune the theory to a critical point, as discussed in Refs. \cite{sub,genem}.}  As such, this theory does not have a natural duality between the electric and magnetic sector.  From its definition, we see that the magnetic tensor obeys the divergence-free condition $\partial_iB^{ij} = 0$.  When the gauge theory is compact, such that $A_{ij}$ is only defined mod $2\pi$, this condition relaxes to $\partial_iB^{ij} = \Tilde{\rho}^j$, where $\Tilde{\rho}^j$ represents the density of vector-flavored magnetic monopoles.  Importantly, however, compactness does not lead to instantons ($i.e.$ flux slip events) in three spatial dimensions, which allows for the existence of a stable deconfined phase \cite{rasmussen}.  (This is in contrast to two-dimensional compact theories, which are destabilized by instantons.)

Using the gauge-invariant fields $E_{ij}$ and $B_{ij}$, we can immediately write down a Maxwell-type Hamiltonian for this symmetric tensor gauge theory as:
\begin{equation}
H = \int d^3x\frac{1}{2}(E^{ij}E_{ij} + B^{ij}B_{ij})
\end{equation}
Note that we have neglected a potential trace term of the form $(E^i_i)^2$, which turns out to be an irrelevant perturbation to this fixed point \cite{sub}.  We have also restricted our attention to rotationally invariant theories, whereas other terms may generically be present in systems with a particular lattice symmetry.  By calculating the equations of motion of this quadratic Hamiltonian, it is a straightforward exercise to show that this model gives rise to five gapless gauge modes with a linear dispersion, $\omega\sim k$.  These gauge modes are simply the natural tensor analogue of photons, which may also be regarded as ``gravitons" in light of our later discussion connecting with gravity.  So far, there have been few surprising aspects to this tensor Maxwell theory, which behaves very much like normal Maxwell theory with a few extra indices.

The unusual aspects of this theory arise when we consider the electric charge sector of the theory.  From the gauge transformation $A_{ij}\rightarrow A_{ij} + \partial_i\partial_j \alpha$, as well as the canonical conjugate relationship between $A_{ij}$ and $E_{ij}$, we can immediately deduce that the electric field obeys the constraint $\partial_i\partial_jE^{ij}=0$ within the pure gauge theory.  (One way to see this is to note that $\partial_i\partial_j E^{ij}$ effectively acts as the generator of the gauge transformation.)  This constraint serves as the source-free Gauss's law of the theory.  Naturally, we can then loosen this constraint by introducing a charge density $\rho$, leading to the full Gauss's law:
\begin{equation}
\partial_i\partial_jE^{ij} = \rho
\end{equation}
which is the single most important equation of the entire theory.  While this tensor Gauss's law may look only mildly different from the familiar vector one, it leads to a dramatic consequence for the mobility of charges, as encoded in the conservation laws of the theory.  As in ordinary Maxwell theory, the Gauss's law immediately dictates that charge is locally conserved.  Formally, one can write the charge within some region of space as:
\begin{equation}
Q = \int d^3x\,\rho = \int d^3x\,\partial_i\partial_j E^{ij} = \oint dn_j\,\partial_iE^{ij}
\end{equation}
where in the final step we have rewritten the integral of a divergence as a flux through the boundary.  This equation tells us that the charge in any region of space can only change via the flux of charge in or out through the boundary.  In other words, charge is a locally conserved quantity in the bulk of the system.  While the conservation of charge is to be expected, this theory also contains a second type of conservation law with no analogue in ordinary Maxwell theory.  Specifically, let us consider the dipole moment associated with the charge in some region of space:
\begin{equation}
P^i = \int d^3x\,\rho x^i = \int d^3x\,x^i\partial_j\partial_kE^{jk} = \oint dn_k\,(x^i\partial_jE^{jk}-E^{ik})
\end{equation}
where we have integrated by parts and taken advantage of divergences to arrive at a boundary term.  We see that, just like charge, the dipole moment of this theory can also be written in terms of a flux through the boundary.  This indicates that dipole moment is also a locally conserved quantity in the bulk of the system.

This local conservation of dipole moment immediately leads to the fracton phenomenology illustrated in Figure \ref{fig:fracton}.  An isolated charge is incapable of moving, since any motion will change the dipole moment of the system.  Meanwhile, a dipolar bound state is free to move, so long as it preserves the magnitude and orientation of its dipole moment.  It is also possible for a single fracton to move alongside the simultaneous creation of an additional dipole, such that the total dipole moment remains invariant.  However, such a process requires a large input of energy to create the new dipole.  To maintain constant motion via this mechanism, a fracton require a constant input of energy in order to create new dipoles at every step.  We therefore conclude that the charges of this symmetric tensor gauge theory are the prototypical example of fracton excitations.

This tensor gauge theory has other important features, such as a full set of tensor Maxwell equations.  We will also see later how this gauge theory draws unexpected connections between fractons and topics such as elasticity and gravity.  For now, however, we conclude this overview of tensor gauge theories by briefly mentioning a second type of theory with slightly different properties.  In addition to the ``scalar charge" tensor gauge theory we have been discussing, it is also possible to write down a ``vector charge" tensor gauge theory governed by a different type of gauge transformation, $A_{ij} \rightarrow A_{ij} + \partial_i\alpha_j + \partial_j\alpha_i$, along with a corresponding vector-flavored Gauss's law:
\begin{equation}
\partial_iE^{ij} = \rho^j
\end{equation}
As in the scalar charge theory, these vector charges will have restricted mobility due to an unusual set of conservation laws.  This theory exhibits both conservation of vector charge, $\Vec{Q} = \int d^3x\,\Vec{\rho}$, and conservation of the angular charge moment, $\Vec{M} = \int d^3x\,(\Vec{\rho}\times\Vec{x})$.  This set of conservation laws leads to the restriction that each vector charge can only move along the direction of its charge vector, while motion in the transverse directions is prohibited.  We therefore refer to the charges of this second type of tensor gauge theory as one-dimensional particles, or lineons.  Through further modifications to the tensor gauge field, $e.g.$ by adding additional indices or imposing tracelessness conditions, it is possible to get other types of particles with restricted mobility, such as two-dimensional particles or fractons exhibiting a conserved quadrupole moment.  While the precise nature of the restricted mobility varies from one theory to another, such restrictions appear to be a generic feature of symmetric tensor gauge theories.

\subsection{Advances in Tensor Gauge Theory}

Having established the basic physical principles of tensor gauge theories, we now provide a brief overview of some recent advances in this area of study.  The casual reader interested mainly in a broad introduction to fractons may wish to only skim this subsection on a first reading.

\subsubsection{Fracton Field Theories}

Throughout this section, we have shown how fractons naturally arise in tensor gauge theories.  In the continuum limit, these gauge fields are governed by a tensor Maxwell theory, serving as a field theory description for the gauge sector.  However, we have so far not discussed how one can write down a field theory for the actual fractons themselves.  To illustrate how this can be done, we will work with the simplest case of a theory obeying conservation of charge and dipole moment.  We will first show how to write a field theory consistent with charge and dipole conservation, then show how that theory can be gauged to yield the scalar charge tensor gauge theory \cite{principle}.  Similar considerations can then be applied to other types of fractons and subdimensional particles, leading to different types of tensor gauge theories.

We start by writing a complex scalar field $\Phi$ to describe fracton matter, and we assume our theory is invariant under global phase rotations, $\Phi\rightarrow e^{i\alpha}\Phi$, corresponding to conservation of charge.  However, we also stipulate that the theory is invariant under linear phase rotations, $\Phi\rightarrow e^{i\vec{\lambda}\cdot\vec{x}}\Phi$ for constant $\vec{\lambda}$, corresponding to conservation of dipole moment \cite{principle}.  To construct a Lagrangian invariant under this transformation, it is useful to first construct covariant operators, transforming only via a phase factor.  In contrast to ordinary field theories, however, this theory does not possess any covariant operators featuring spatial derivatives acting on only a single $\Phi$ operator.  Rather, the lowest order covariant spatial derivative operator contains two factors of $\Phi$, taking the form:
\begin{equation}
\Phi\partial_i\partial_j\Phi - \partial_i\Phi\partial_j\Phi
\end{equation}
which can be checked to transform covariantly.  Using the covariant operators, we can then write down a lowest order Lagrangian for this theory as:
\begin{equation}
\mathcal{L} = |\partial_t\Phi|^2 - m^2|\Phi|^2 - g |\Phi\partial_i\partial_j\Phi - \partial_i\Phi\partial_j\Phi|^2 - g' \Phi^{*2}(\Phi\partial_i\partial_j\Phi - \partial_i\Phi\partial_j\Phi)
\end{equation}
which takes a characteristically non-Gaussian form.  (Such a non-Gaussian field theory was also encountered earlier in a field-theoretic treatment of the X-cube model \cite{Slagle2017-gk}.)  Very little is currently known about the properties of this field theory.  Can one explicitly calculate the correlators of this model, perhaps via a perturbative diagrammatic method?  Can the dipole dispersion be directly extracted from the Lagrangian?  Is there some useful renormalization group scheme which can be applied to this theory?  These all remain interesting open questions.  At present, the one thing which is known about this theory is how to gauge it.  Let us now stipulate that our theory must be invariant under phase rotations with arbitrary spacetime dependence, $\Phi\rightarrow e^{i\alpha(x,t)}\Phi$.  Under such a transformation, our previously covariant operator transforms as $\Phi\partial_i\partial_j\Phi - \partial_i\Phi\partial_j\Phi \rightarrow e^{2i\alpha}[\Phi\partial_i\partial_j\Phi - \partial_i\Phi\partial_j\Phi +(i\partial_i\partial_j\alpha)\Phi^2]$.  We can then construct a gauge-covariant derivative operator by introducing a gauge field $A_{ij}$ transforming as $A_{ij}\rightarrow A_{ij} + \partial_i\partial_j\alpha$, which enters the derivative operator as:
\begin{equation}
\Phi\partial_i\partial_j\Phi - \partial_i\Phi\partial_j\Phi - iA_{ij}\Phi^2
\end{equation}
We see that the gauge tranformation of the tensor gauge field is precisely that of the scalar charge theory, and indeed this gauge-covariant derivative can be used to write down a field theory describing both the matter and gauge sectors of the scalar charge theory.

A natural generalization of this scalar chage theory is developed in Ref.~\cite{bulmash2018generalized} with a
class of generalized $U(1)$ gauge theories whose charge excitations exhibit fractal structure akin to type-II fracton models \cite{haah}. We note that there have also been many other recent developments regarding fractons and field theories, and we refer the interested reader to the literature for details \cite{curved,mikeleo,nati,smn,juven1,juven2,radi,structure}.

\subsubsection{Generalized $U(1)$ Symmetry and the Multipole Algebra}

Motivated by the aforementioned fracton gauge principle perspective \cite{principle}, a generalized fracton theory can be acquired by gauging a charged matter field with generalized $U(1)$ symmetry and conserved multipole moment. This approach can be systematically tackled in terms of the notion of a multipole algebra which is a natural generalization of the symmetry algebras generated by the polynomial shift symmetries in Ref.~\cite{shift}.

In Ref.~\cite{gromov2018towards}, the author demonstrated that upon gauging the generalized $U(1)$ symmetry one finds the symmetric tensor gauge theories, as well as the generalized gauge theories discussed recently in the literature \cite{bulmash2018generalized}.  The outcome of the gauging procedure depends on the choice of the multipole algebra. Such generalized $U(1)$ symmetries with conserved multipole moment cannot be regarded as "internal" because they do not commute with spatial translations and rotations. Upon a unique gauging procedure with proper UV regularization depending on the choice of the multipole algebra, one eventually reaches symmetric tensor gauge theories akin to the recently discussed generalized gauge theories.

On a parallel and alternative search, one can show that fractonic matter naturally appears in vector gauge theories enriched by global $U(1)$ and translational symmetries, via the mechanism of `anyonic spin-orbital' coupling \cite{seu1}.  Namely, if the global symmetry quantum number is changed upon the translation of a quasiparticle, then moving the charged particles out of the submanifold is clearly forbidden when the global symmetry is present. If the global symmetry is then gauged, the restricted particles become fractons as moving a fracton breaks gauge invariance. More generally, the relation between symmetry restrictions on the mobility of quasiparticles and symmetry-enriched topological orders relies on the fact that the actions of translation and global symmetries on quasiparticle excitations do not commute. This line of thinking opens a new page to connect fracton phases of matter and spatial symmetry enriched topological ordered phases and identifies new specimens of fractonic matter in these settings.

\subsubsection{Tensor Chern-Simons Theories}

Given the power of Chern-Simons gauge theory to study topological orders in 2 dimensions, it is natural to ask whether there is a class of fractonic Chern-Simons theories which capture fractonic behavior.  Clearly, such field theories must be both similar to, and qualitatively different from, TQFTs---in which the details of the underlying lattice (or regularization) are unimportant, and universal topological physics emerges.  A number of possible approaches to this challenge have been discussed in the literature thus far from the spirit of BF-type theory \cite{Slagle2017-ne,hanlayer,Slagle2017-gk,witten,higgs1,prem2017emergent,sub,higgs2,bulmash2018generalized,you2018symmetric,gromov2018towards,gromov}. These BF theories, with proper lattice regularization, can be viewed as the $U(1)$ limit of all CSS quantum stabilizer codes.

However, other non-CSS stabilizer codes, such as the Chamon code \cite{chamon}, do not admit such a BF description.  In Ref. \cite{you2019fractonic}, the authors proposed a lattice version of the fractonic Chern-Simons theory inspired by the spirit of flux attachment. By imposing a constraint binding charge to the flux of a higher-rank gauge field, the fractonic gauge flux is decorated with a gauge charge with similar subdimensional mobility. Such a fractonic flux attachment procedure introduces a non-commutative gauge structure and thus creates a deconfined $U(1)$ fracton theory. Although such fractonic Chern-Simons theories are clearly not TQFTs, they share several important features of the {\it chiral} 2+1D Chern-Simons theories.  First, the fractonic Chern-Simons term creates self-statistical interactions between charged excitations.  Second, the fractonic Chern-Simons action is gauge invariant only up to a boundary term,  implying that their boundaries host gapless surface states that cannot be realized in 2 dimensions with subsystem symmetry.  These are closely related to the surface states of subsystem-symmetry protected models described in Ref. \cite{you2018symmetric}.  We also note that chiral two-dimensional tensor Chern-Simons theories can occur at the boundary of certain three-dimensional fracton phases \cite{witten}.

The starting point to construct a tensor Chern-Simons term is to seek a symmetry structure with 2 spatial gauge fields $A_1$ and $A_2$, which will allow us to obtain a fully gapped Chern-Simons theory with a single constraint.  Consider gauge transformations of the form
\be 
A_1 \rightarrow A_1 + D_1 \alpha  \ , \ \ A_2 \rightarrow A_2 + D_2 \alpha \ 
\ee
where $ D_1$ and $D_2$ are differential operators, whose form we will leave unspecified for now.   
Since we only have 2 gauge fields, the magnetic field defined has a single component\footnote{Note that the magnetic field (\ref{Eq:Bfield}) is always gauge invariant; however it is not necessarily the most relevant gauge invariant magnetic field that we can write down.  If $D_1$ and $D_2$ share a common factor $\partial_\ell$, the operator $\partial_{\ell}^{-1} B$ is also gauge invariant. }
\be \label{Eq:Bfield}
B = D_2 A_1 - D_1 A_2 \ .
\ee
The gauge-invariant electric fields have the form
\be \label{Eq:Efield}
E_i = \partial_t A_i - D_i A_0
\ee
where we have introduced the usual time component of the gauge field, which transforms as
\be
A_0 \rightarrow A_0 + \partial_t \alpha
\ee
under gauge transformations.

The generalized Chern-Simons action has the form,
\be \label{LCS0}
\mac{L}_{CS} = \frac{s}{4 \pi} \left( A_1 E_2 - A_2 E_1 - (-1)^\eta A_0 B \right)
\ee
where $\eta = 1$ if the $D_i$ contain only even numbers of derivatives, and $\eta = 2$ if they contain only odd numbers of derivatives.  Under gauge transformations, 
\ba  \label{Eq:CSGtrans}
\delta \mac{L}_{CS} &=& 
\frac{s}{4 \pi} ( D_1 \alpha E_2 - D_2 \alpha E_1 - (-1)^\eta \partial_t \alpha B )\n
&=&\frac{s}{4 \pi} (  D_1 \alpha \partial_t A_2 + (-1)^\eta \partial_t \alpha D_1 A_2 \n
&& -( D_2 \alpha \partial_t A_1 + (-1)^\eta \partial_t \alpha D_2 A_1) \n
&& +  D_2 \alpha D_1 A_0 -  D_1 \alpha D_2 A_0 )
\ea
In the absence of boundaries, one may freely integrate by parts, to obtain:
\be
\delta \mac{L}_{CS; \text{Bulk}} = 
0
\ee
The boundary terms in general do not vanish, implying the existence of gapless boundary modes, whose precise nature depends on the choice of $D_i$.  

Irrespective of the choice of $D_i$, the Chern-Simons action (\ref{LCS0}) has several commonalities with the standard vector Chern-Simons theory in $2+1$ dimensions.  First, in the absence of sources the constraint simply sets $B=0$.  Since there is only one component of the magnetic field, this one constraint is sufficient to eliminate the possibility of any propagating gauge degrees of freedom, leading to a gapped theory whose physics is entirely determined by operators describing pure gauge degrees of freedom.\footnote{This only applies to the case where $D_1,D_2$ do not share any common factor. Otherwise, even though the magnetic flux fluctuation is fixed, there might exist some local operator with lower order exhibiting a dispersive gapless mode.}  In ordinary Chern-Simons theory these are the holonomies, or gauge-invariant Wilson lines along non-contractible curves.  We will discuss the analogue of Wilson line operators  for specific examples of $D_i$ in detail presently; these have the general form $e^{i\int_s A_i }$ with the submanifold $s$ chosen to ensure the operator is gauge invariant.

Second,  irrespective of the choice of $D_i$, the gauge fields $A_1$ and $A_2$ are canonically conjugate.  If both gauge fields are compact, this implies that a generalized Wilson operator of the form $e^{i\int_s A_i }$ must be discrete as well as compact.  Thus each of the generalized Wilson operators can take on only a finite, discrete set of values, which fully specify the states allowed in the absence of sources.  On closed manifolds this can give either a finite or a countable ground state degeneracy. 

Finally, in the presence of matter fields, the Chern-Simons action (\ref{LCS0}) has the effect of binding charge to flux.  To see this, we add matter fields to our Chern-Simons action in the standard way, by adding a term
\be
\mac{L}_{\text{Matter}} = A_0 \rho - A_i J^i
\ee
where the currents obey the conservation law:
\be
D_i J^i = \partial_t \rho
\ee
Depending on the specific form of the differential operator $D_i$, the theory might contain additional subsystem charge conservation laws and charge multipole conservation \cite{gromov2018towards}.  In the presence of sources the Chern-Simons constraint is
\be
B = D_2 A_1 - D_1 A_2 = \frac{2 \pi}{s} \rho
\ee
which binds the generalized magnetic flux to charge.  One might anticipate that a generalized Aharonov-Bohm effect may endow these charge-flux bound states with fractional statistics.  Indeed, as gauge invariant operators involving $A_1$ do not commute with gauge-invariant operators involving $A_2$, we will usually find at least some excitations with nontrivial mutual statistics.

\subsubsection{Generalized Witten Effect}

In Maxwell theory, the axion term $\theta \vec{E} \cdot \vec{B}$ is a total derivative which has no effect on the gapless photon, but has two important, closely related consequences: attaching electric charge to magnetic monopoles (the Witten effect) and leading to a Chern-Simons theory on the boundary. A similar story\cite{witten} holds in the higher rank $U(1)$ gauge theories which admit generalized axion terms which intertwine higher rank electric field with tensor gauge flux. Such axion terms
have no effect on the gapless gauge mode, but bind together electric and magnetic charges (both of which are generally subdimensional) in specific combinations, in a manifestation of the Witten effect. In particular, the axion term could have quantized $\theta$ value provided time-reversal invariance is imposed. In addition, these axion terms in tensor $U(1)$ gauge field imply a non-trivial boundary structure with a Chern-Simons-like action in both chiral and non-chiral settings. 

The search for topological $\theta$ terms in fractonic phases of matter can generate a rich sequence of new fractionalized fracton theories whose intrinsic features still remain to be unlocked. For example, the fractonic $\theta$ term also exists in 2D characterizing a topological quadrupolarization with fractional corner charge. On the more down-to-earth side of things, it would be highly useful to find concrete lattice models which demonstrate the properties of these $\theta$ terms explicitly as a complement to the field-theoretic approach. In addition, more investigation is required regarding how to measure the topological $\theta$ coefficient or fractonic Witten effect, which will be important for experimental detection of these fracton phases. The axion electrodynamics in tensor gauge theory suggest various directions and open questions for future study. An interesting corrolary of the fractonic Witten effect is that the $\theta$ term in tensor gauge theory can also delineate a topological multipole moment or quantized dipolar Hall effect which characterize a rich class of higher order topological phases\cite{you2019multipolar}. The field theory and topological implications of fractonic axion electrodynamics still remain unclear and thus are worth pursuing further.

\section{Fractons in Solvable Spin Models}
\label{sec:spin}

Some of the most important fracton models were discovered in exactly solvable lattice spin models as quantum error-correcting codes.  In this section, we review some of the paradigmatic examples of solvable spin models exhibiting fracton behavior, such as the X-cube model and Haah's code, along with their relationship to tensor gauge theories.  We then give an overview of more recent developments in the area of fracton spin models, such as various geometric constructions.  We conclude this section by discussing progress towards constructing more realistic spin models for fractons, such as a proposed fracton spin ice.

\subsection{Prototypical Examples}

Two of the most representative fracton spin models are the X-cube model and Haah's code, which are the prototypical examples of type-I and type-II fracton models, respectively.  We review the basic features of each of these models in turn.  Some of the basic properties of the X-cube model and Haah's code are summarized in Table \ref{tab:xc_cc} for comparison.

\begin{table}[h]
    \centering
    \begin{tabular}{| c | c | c |}
    \hline
         & X-cube Model & Haah's Code \\  \hline  \hline
     $\log($GSD$)$    & $2L_x +2L_y+2L_z-3$  & fluctuating, upper bounded by $4L$\\
     \hline
     Fractional excitations & fractons, lineons, planons & fractons only \\
     \hline
     Logical operators & string and membrane & no string, all fractal shaped \\
     \hline
     Sub-region entanglement entropy & Area law $+$ linear correction & Area law $+$ linear correction \\
     \hline
    \end{tabular}
    \caption{Basic properties of the X-cube model and Haah's code. `GSD' stands for ground state degeneracy. The system size for the X-cube model is taken to be $L_x \times L_y \times L_z$ and for Haah's code $L\times L \times L$. }
    \label{tab:xc_cc}
\end{table}

\subsubsection{Type-I Fracton Model: X-cube}

\begin{figure}[htbp]
    \includegraphics[width=0.5\columnwidth]{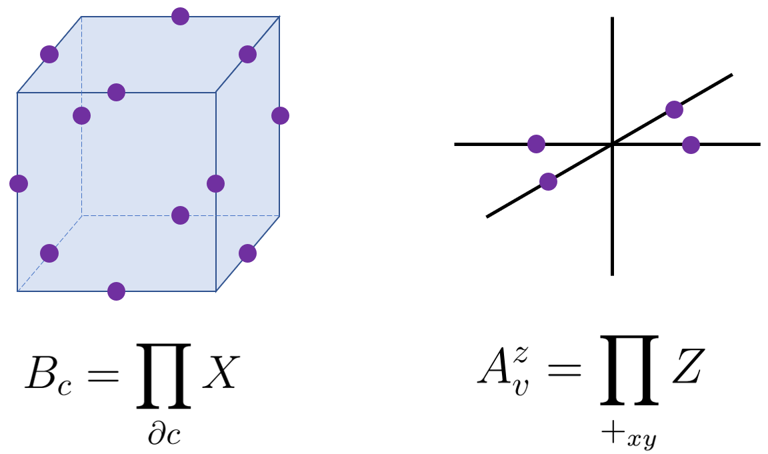}
    \caption{{\bf (a)} A cube operator of the X-cube model is a product of $X$ operators of 12 spins on the edges of a cube;  {\bf (b)} A cross operator is a product of $Z$ operators of 4 coplanar spins touching a vertex.}
    \label{Xc-T3-H}
\end{figure}

The X-cube model, as first discussed in Ref. \cite{vhf2}, is defined on a cubic lattice with qubit degrees of freedom on the edges.  The Hamiltonian
\begin{equation}
H = -\sum_v \left(A_v^{x}+A_v^{y}+A_v^{z}\right) -\sum_c B_c   
\label{eq:H}
\end{equation}
contains two types of terms (\figref{Xc-T3-H}): cube terms $B_c$ which are products of the twelve Pauli $X$ operators around a cube $c$, and cross terms $A^{\mu}_v$ which are products of the four Pauli $Z$ operators at a vertex $v$ in the plane normal to the $\mu$-direction where $\mu=x,y,\text{ or }z$.  These terms mutually commute and their energies can be minimized simultaneously. 

On a $L_x\times L_y\times L_z$ cubic lattice with periodic boundary conditions, the log of the ground state degeneracy (GSD) scales linearly with the size of the system in all three directions:
\begin{equation}
    \log_2{\textrm{GSD}}=2L_x+2L_y+2L_z-3.
    \label{eq:GSDtorus}
\end{equation}

\begin{figure}[htbp]
    \centering
    \includegraphics[width=0.8\columnwidth]{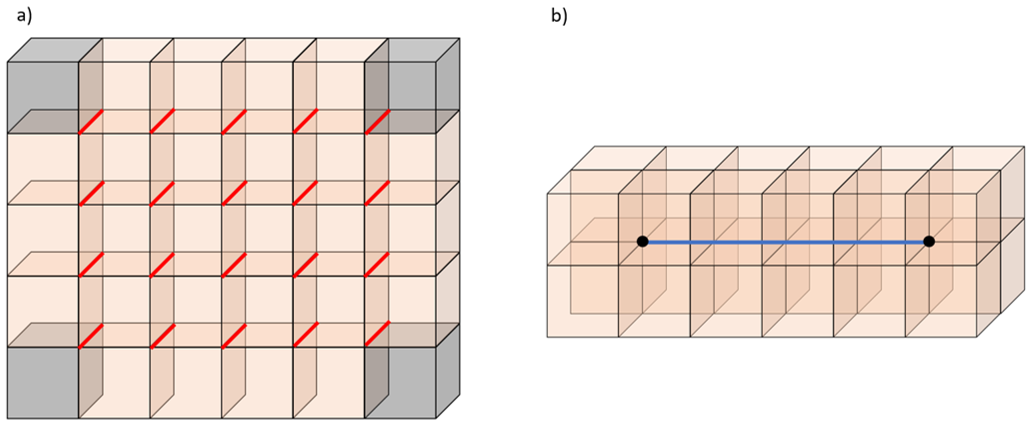}
    \caption{Visualization of particle creation operators.  a) The red links correspond to a membrane geometry on the dual lattice. The product of $Z$ operators over these edges excites four fractons (the darkened cube operators at the corners); b) The product of $X$ operators over the links comprising the straight open blue string creates two lineon excitations at its endpoints (black dots).}
    \label{fig:excitations}
\end{figure}

Fractional excitations can be made by applying string and membrane operators. A product of $Z$ operators over links on a rectangular membrane geometry on the dual lattice anti-commutes with the cube Hamiltonian terms at its corners. See \figref{fig:excitations} a). Applying such a membrane operator hence generates four cube excitations at a time and individually the cube excitations cannot move, forming a `fracton' excitation. A pair of such fracton excitations at adjacent corners may be viewed as a single dipole-like object which is itself a dimension-2 particle and is mobile in the plane normal to the edges connecting the two corners.

A product of $X$ operators over links along a straight line anti-commutes with vertex Hamiltonian terms at the endpoints (\figref{fig:excitations} b)). The vertex excitations are hence created in pairs and can be separated using string operators. But their motion is restricted to one direction only, because the $X$ string operator in different directions anti-commute with a different set of vertex terms, hence creating different excitations. They are called the `lineon' or dimension-1 particles. A pair of lineons separated in the $x$, $y$ or $z$ direction is a dimension-2 particle and is mobile in the plane normal to the edges connecting the two lineons.

In the ground state, the entanglement entropy of a sub-region also contains a linear term. That is, if we take out a sub-region, say of size $R\times R \times R$, and calculate its entanglement entropy, we would find an area law term which scales as $R^2$ and a sub-leading linear term which scales as $R$~\cite{ShiEntropy,HermeleEntropy,SchmitzEntropy,BernevigEntropy}.  (One must take care to avoid any potential spurious contributions to the entanglement entropy, however \cite{williamson2019spurious}.)

\subsubsection{Type-II Fracton Model: Haah's Code}

\begin{figure}[htbp]
    \centering
    \includegraphics[width=.7\columnwidth]{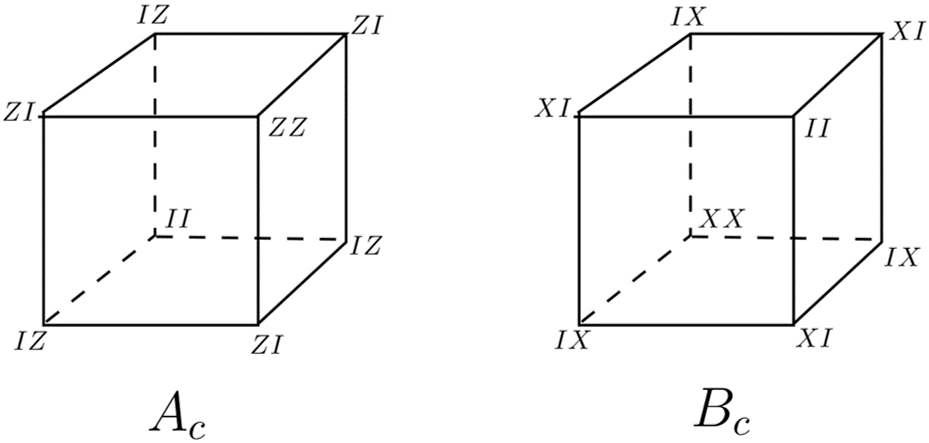}
    \caption{The Hamiltonian of Haah's code is a sum of two types of cube terms.  Recall that there are two qubits per vertex, so $ZZ$ represents a Pauli $Z$ acting on each of the two qubits, for example.  (Figure adapted from Ref. \cite{haah}.)}
    \label{fig:haah}
\end{figure}

\begin{figure}[htbp]
    \centering
    \includegraphics[width=.3\columnwidth]{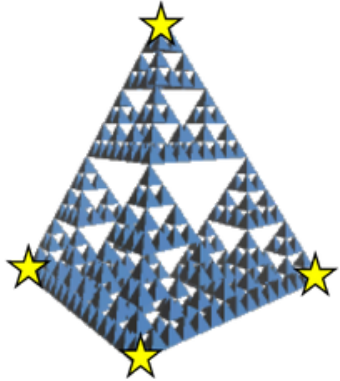}
    \caption{In Haah's code, fractons are created at the corners of fractal operators.}
    \label{fig:fractal}
\end{figure}

We now turn to the paradigmatic example of a type-II fracton model, constructed by Haah in Ref. \cite{haah}, which has since come to be known as Haah's code.  This model is defined on a cubic lattice, with two qubits on every vertex of the lattice.  The Hamiltonian can be written as the sum of two commuting types of cube terms:
\begin{equation}
H = -\sum_c A_c -\sum_c B_c
\end{equation}
where $A_c$ is a particular product of $Z$ operators touching a cube and $B_c$ is a similar product of $X$ operators, as defined pictorially in Figure \ref{fig:haah}.  Unlike the X-cube model, Haah's code possesses a self-duality between the two types of cube terms, so it is sufficient to study either sector of the theory to obtain a full understanding of its excitation spectrum.  In either sector, application of a single spin operator creates four quasiparticles at the corners of a tetrahedron.  Repeated application of spin operators in a specific pattern can separate these particles to the four corners of a fractal operator, as indicated schematically in Figure \ref{fig:fractal}.  However, there is no string operator which can move these particles individually around the system, so these are immobile fractons.  It can further be proved that there are no string logical operators in the theory whatsoever \cite{haah}, indicating that all nontrivial bound states of the fractons are also immobile, making this a type-II fracton model.

Like the X-cube model, Haah's code also exhibits a subextensive ground state degeneracy
, albeit with a more complicated dependence on system size.  For a 3-torus of size $L\times L\times L$, the ground state degeneracy is upper-bounded by $\log_2 GSD < 4L$.  However, at certain special system sizes, the degeneracy can be far less \cite{haah}.  In contrast, the entanglement entropy of Haah's code has a much simpler dependence on subsystem size.  For a subsystem of linear size $R$, the entanglement entropy obeys an area law with a linear subleading correction, just as for the X-cube model \cite{HermeleEntropy}.

\subsection{Higgsing}

The spin fracton models are very different from the $U(1)$ tensor gauge theories. They are gapped and formulated as lattice models rather than field theories. On the other hand, they share the crucial property of hosting fractional excitations with restricted motion. A natural question to ask is whether they are related in some ways. For example, could the spin model be a `Higgsed' version of the $U(1)$ tensor gauge theory such that only a discrete subgroup of $U(1)$ is preserved? It was found that, this is indeed the case sometimes, but whether or not a $U(1)$ tensor gauge theory gives rise to a fracton spin model upon Higgsing depends sensitively on the form of the conservation law of the $U(1)$ theory \cite{higgs1,higgs2}. For example, the scalar charge theory becomes non-fractonic once Higgsed while a modified `hollow' tensor gauge theory remains fractonic even upon Higgsing. For the following discussion, we take, WLOG, the Higgsed gauge group to be $\mathbb{Z}_2$ and the gauge charges live on the cubic lattice sites ${\bf r}=(x,y,z)$, with lattice spacing $a=1$.

The result of Higgsing can be directly studied through its effect on the conservation laws: charge conservation, dipole conservation, etc. Suppose we start with a $U(1)$ gauge theory with charge conservation. That is the total charge $Q$ in a region cannot be changed by acting with local operators within that region. Upon Higgsing the theory to $\mathbb{Z}_2$, charge-$2$ objects can appear from and be absorbed into the condensate.  Therefore $Q$ is now well-defined only modulo $2$, but the conservation of $\mathbb{Z}_2$ charge puts no constraints on the mobility of charges.

\begin{figure}[htbp]
    \centering
    \includegraphics[width=.7\columnwidth]{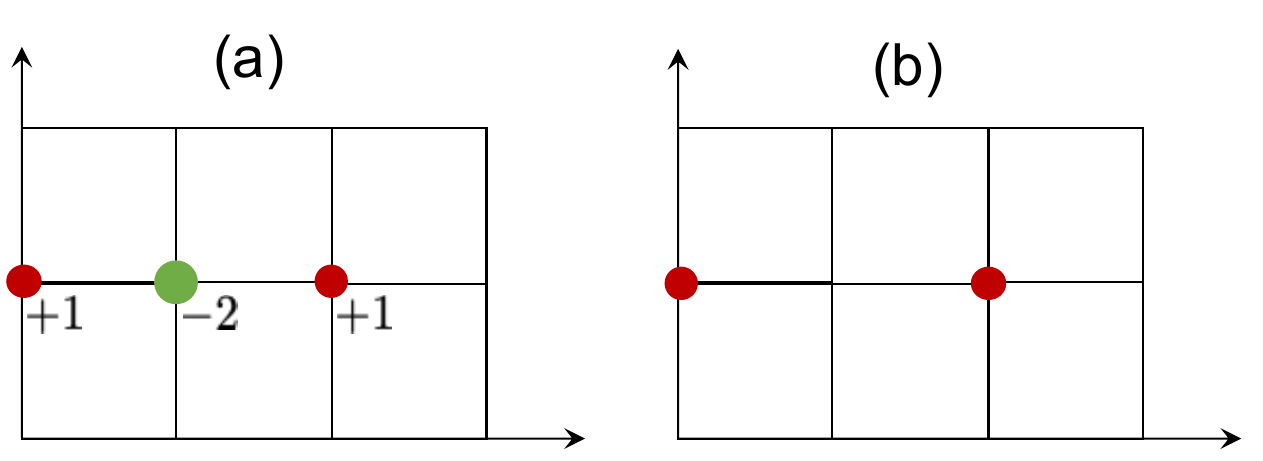}
    \caption{Local $U(1)$ charge configuration with zero total charge and zero total dipole (a) reduces to charge pair creation / hopping; (b) when the $U(1)$ gauge theory is Higgsed to $\mathbb{Z}_2$.}
    \label{fig:dipole}
\end{figure}

What about the conservation of dipole moment? In the $U(1)$ theory, the dipole moment $\boldsymbol{d}$ of some region V is given by $\boldsymbol{d} = \sum_{{\bf r} \in V} {\bf r} \, n_{\bf r}$, and is conserved in the sense that it cannot be changed locally. Higgsing to $\mathbb{Z}_2$ simply means that $\boldsymbol{d} \mod N$ is conserved, \emph{i.e.} each component of $\boldsymbol{d}$ is separately conserved modulo $N$. As we set the lattice constant to one; then charge $\pm 2$ objects appearing from the condensate can change each component of $\boldsymbol{d}$ by integer multiples of $2$.

To understand the effects of the Higgsed dipole conservation law, it is useful to consider locally creatable charge configurations. With both charge conservation and dipole conservation, the configuration shown in \figref{fig:dipole} (a) can be locally created in a $U(1)$ theory. Upon Higgsing, it reduces to pair creation / hopping of $\mathbb{Z}_2$ charges. Therefore, upon Higgsing, the dipole conservation law no longer localizes the gauge charges. The gauge charges can hop in any direction, albeit at distance two at a time. Similar conclusions can be drawn for general $\mathbb{Z}_N$ gauge theories and for general lattice. Therefore, the scalar charge theory become non-fractonic upon Higgsing.

To retain the fractonic nature of the $U(1)$ gauge theory upon Higgsing, stronger conservation laws are needed, for example planar conservation laws. If the charge on every lattice plane is separately conserved, then clearly single charges will not be able to move. This is the case for the `hollow' ${\rm U}(1)$ gauge field with gauge components $E_{xy}, E_{yz}, E_{zx}$ but not the diagonal ones. The Gauss's law is given by
\begin{equation}
\Delta_x \Delta_y E_{xy} + \Delta_y \Delta_z E_{yz} + \Delta_{x} \Delta_z E_{x z} = \rho_{\bf r} \text{,}
\end{equation}
which implies that the total charge on every $x$, $y$, $z$ planes are conserved. Upon Higgsing, it becomes the X-cube model discussed in the previous section.

\subsection{Geometric Aspects}

Unlike spin models with topological order, such as Toric Code, the fracton spin models seem to care not just about the topology of the underlying manifold, but also the geometry of the underlying lattice. For example, in \Ref{Slagle2017-gk}, it was noticed that spatial curvature can induce a stable ground state degeneracy for the X-cube model. In the following sections, we review the coupled layer approach and the cage net and string-membrane net approach which not only help to elucidate the geometric nature of these models, but also lead to the construction of new models.

\subsubsection{Coupled Layer Constructions} 
\label{sec:CL}

\begin{figure}[htbp]
    \centering
    \includegraphics[width=1.0\columnwidth]{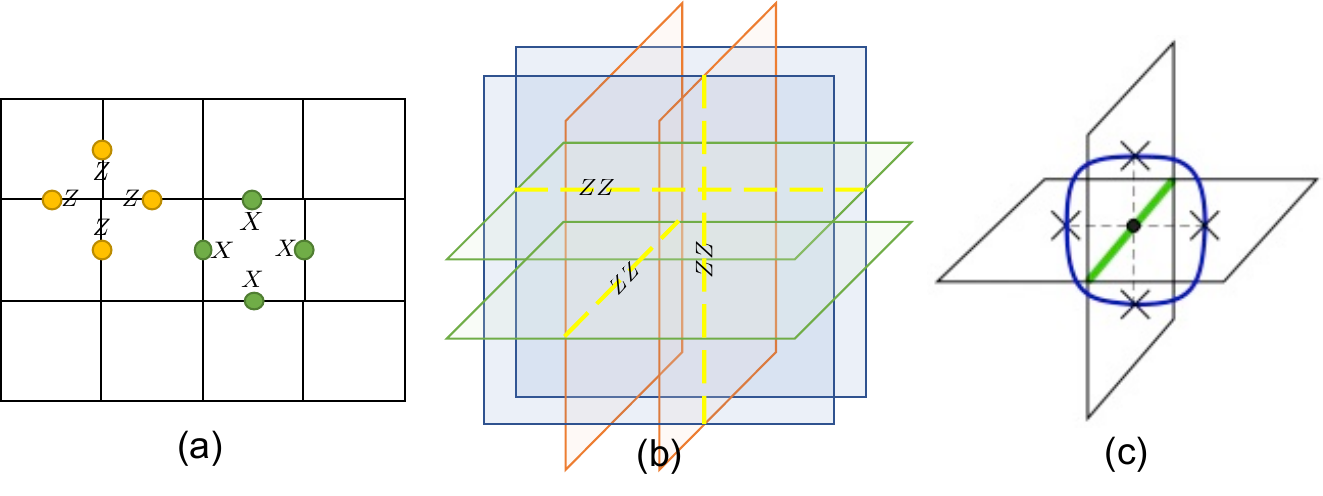}
    \caption{The coupled layer construction: (a) take 2D Toric Code model; (b) make three intersecting stacks of them and couple strongly along the intersection line; (c) the coupling generates flux particle loops which are condensed in the strongly coupled limit.}
    \label{fig:CL}
\end{figure}

In \Ref{hanlayer,sagarlayer}, it was noticed that the X-cube model can be obtained by taking the 2D Toric code model (\figref{fig:CL} (a)), make three intersecting stacks in $xy$, $yz$ and $zx$ planes respectively, and couple them along intersection lines where the edges overlap with a $Z\otimes Z$ coupling term (\figref{fig:CL} (b)). When the coupling terms become large, the coupled model becomes effectively the X-cube model. 

It was found that the coupling process induces a `particle-loop' condensation, driving the phase transition from a stack of 2D topological order to a 3D fracton order. In particular, the $Z\times Z$ term creates one pair of flux particle in each of the intersecting planes. Taken together, the four flux particles connect into a small loop. When the coupling term becomes large, such small loops condense. The ground state wave function becomes a superposition of `particle-loops' of all shapes and sizes. 

The change in quasi-particle type follows accordingly. As the flux particle loops now form a condensate, individual flux particles are no longer excitations as long as they form a loop. Instead, the end points of flux particle loops become excitations. As the flux particles always appear in pairs, the end of particle loops always appear in a set of four and individually they cannot move, hence becoming the fractons. On the other hand, individual charge particles of the 2D topological order get confined because of its nontrivial statistics with the condensate. A bound state of charge particles on intersecting planes remain deconfined, but can only move along the intersection line of the two planes, hence becoming the lineon.

We note that recent progress has also been reported on coupled-layer constructions for type-II fracton models \cite{schmitz2019distilling}.

\subsubsection{Cage-Net Models}

\begin{figure}[htbp]
    \centering
    \includegraphics[width=0.8\columnwidth]{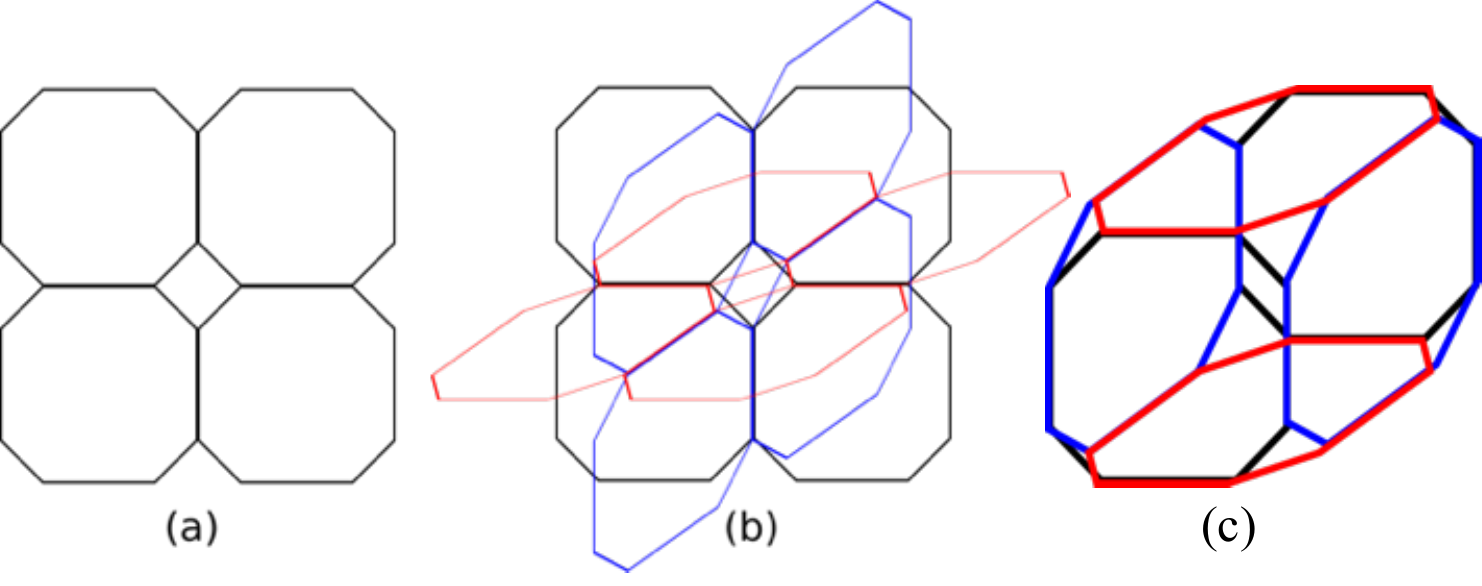}
    \caption{The cage-net models: (a) take 2D string-net states on a decorated (trivalent) square lattice; (b) make three intersecting stacks of them and couple strongly along the intersection line; (c) the resulting wave-function is a superposition of cage-nets configurations, the simplest of which is shown here.}
    \label{fig:CN}
\end{figure}

The cage-net construction \cite{cage} generalizes the coupled layer construction discussed above from stacks of Toric code to stacks of other string-net states. To accommodate more interesting string-net states, the square lattice in each layer is replaced by a tri-valent modification of it (\figref{fig:CN} (a)). When the intersecting stacks are put together, a similar coupling term is added to condense particle loops. As a result, some 2D fractional excitations further fractionalize into fractons while others (in intersecting perpendicular planes) bind into lineons.

The name `cage-net' generalizes the idea of `string-net'\cite{string-net} for 2D topological orders. In 2D, if the DOF in a certain basis are interpreted as representing different string types, then a `string-net' wave function is a superposition of all (branching) loop configurations on a 2D graph satisfying a set of conditions. `String-net' is a systematic way for constructing exactly solvable lattice models for 2D non-chiral topological orders. In `cage-net', as the strings on perpendicular planes are bound together by the coupling term, they form `cage'-like shapes, as shown in \figref{fig:CN}. The ground state wave function is a superposition of all (branching) cage configurations satisfying a set of conditions and host fracton order.

One of the most interesting `cage-net' models is obtained by stacking 2D doubled Ising models and couple them by binding the Ising strings on intersecting planes. The resulting model has fracton and lineon excitations, and the interesting feature is that the lineon is non-abelian. The non-abelian-ness of this model is intrinsically 3D, as it was shown in \Ref{cage} it cannot come from nonabelian 2D models inserted into an otherwise abelian fracton model. 

\subsubsection{String-Membrane-Net Construction}

In \Ref{smn}, a string-membrane-net picture was proposed and was shown to construct models which are equivalent to most of the known foliated fracton models (discussed in the next section) up to trivial degrees of freedom and local unitary transformations. 

\begin{figure}[htbp]
    \centering
    \includegraphics[width=1.0\columnwidth]{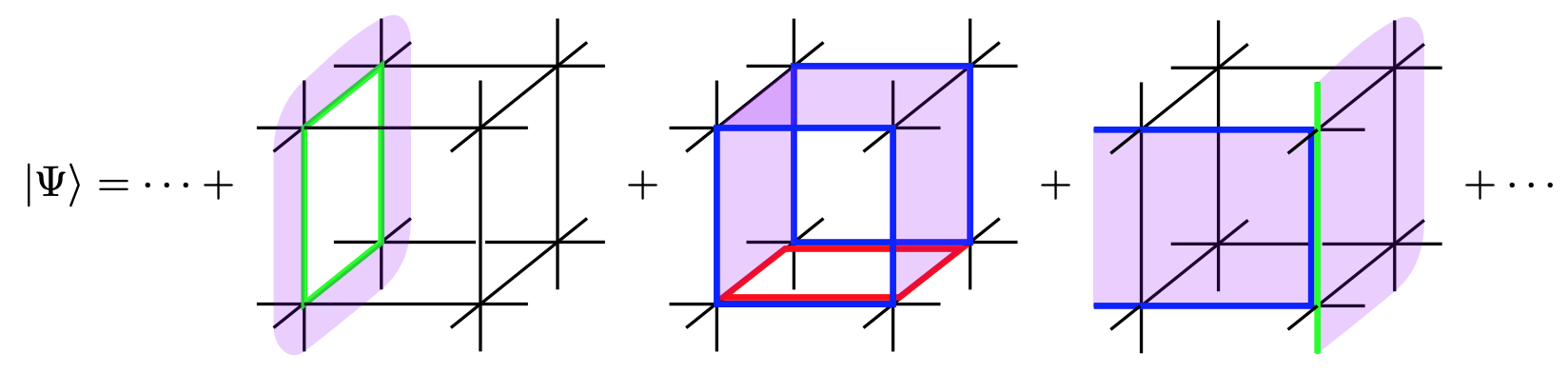}
    \caption{Ground state wave-function of string-membrane-net models as a superposition of all allowable string-membrane-net configurations. Purple plaquettes belong to the membranes and the red, blue, green edges belong to the string-nets in the $xy$, $yz$, $zx$ planes respectively. The constraint is that the edge of the membrane has to be attached to strings.}
    \label{fig:smn}
\end{figure}

The idea is to have two sets of degrees of freedom, one on the plaquettes of a 3D lattice (e.g. a cubic lattice) and the other on the edges of 2D lattices on sets of 2D layers (e.g. square lattice on $xy$, $yz$, and $zx$ planes). The 3D lattice and 2D lattices are arranged such that the edges of the 3D lattice overlaps with edges on the 2D lattices on intersecting planes. The degrees of freedom on the 2D layers form string-nets, as prescribed in \Ref{string-net}. The degrees of freedom on the 3D lattice form membranes. Moreover, they are coupled such that the edges of the 3D membrane is attached to 2D strings. The ground state wave-function is then a superposition of all such `string-membrane-net' configurations, as shown in \figref{fig:smn}

A local Hamiltonian can be written down to have this wave-function as the gapped ground state. It was shown that by changing the number of sets of 2D layers and the type of string-net or the type of membranes, it is possible to construct models which are equivalent to a variety of known fracton models, including the X-cube model and its $\mathbb{Z}_N$ generalization, stacks of Toric Code, and the lineon model discussed in \Ref{fol3}. A particularly nice feature of the string-membrane-net construction is that it readily gives rise to a field theory description of the constructed models. In particular, a very important feature of such models -- the foliation structure discussed in more detail in \secref{sec:foliation} -- shows up nicely in this construction and allows natural field theory representation.

\subsection{Towards Realistic Spin Models}

Despite the exact-solvability of fracton models in quantum stabilizer codes, most of these models require complicated spin-cluster interactions, which seems superficial and unreachable in real materials or cold atom system. To conquer the complexity of the aforementioned fracton codes, the authors in Ref.~\cite{Slagle2017-ne} introduced and proposed a series of quantum spin models in frustrated magnetism which only involves nearest-neighbor two-spin interactions. Nevertheless, these frustrated spin systems, despite lack of exact solvability, still exhibits a stable fracton phase. 

In Ref.~\cite{halasz2017fracton}, the authors proposed a systematic route to construct realistic spin models hosting fracton phases in terms of strongly coupled spin chains. Such coupled spin chain constructions merely require stacking of 1d spin model with  spin bilinear inter-chain interactions, which is more amenable to a potential experimental implementation. Following this spirit, Slagle and Kim~\cite{Slagle2017-ne} proposed a Kitaev type three-dimensional hyper honeycomb lattice with frustrated spin bilinear coupling in different directional bonds. Amazingly, this model, in different limits, displays either 3D fracton order or supports a $Z_2$ spin liquid phase. Thanks to the rapid development of the Kitaev materials discovered in correlated spin orbital coupled system including $Na_2IrO_3$, $\alpha$-$Li_2IrO_3$, $\alpha$-$RuCl_3$ and $H_3LiIr_2O_6$, we expect there may exist a material candidate for such a fracton phase.  There have also been recent simplifications to the Slagle-Kim construction which are promising for material realization \cite{fuji}.  Additionally, synthetic quantum matter, such as AMO experiments, provide another possible route to the experimental realization of fracton phases.

\begin{figure}[htbp]
    \centering
    \includegraphics[width=0.5\columnwidth]{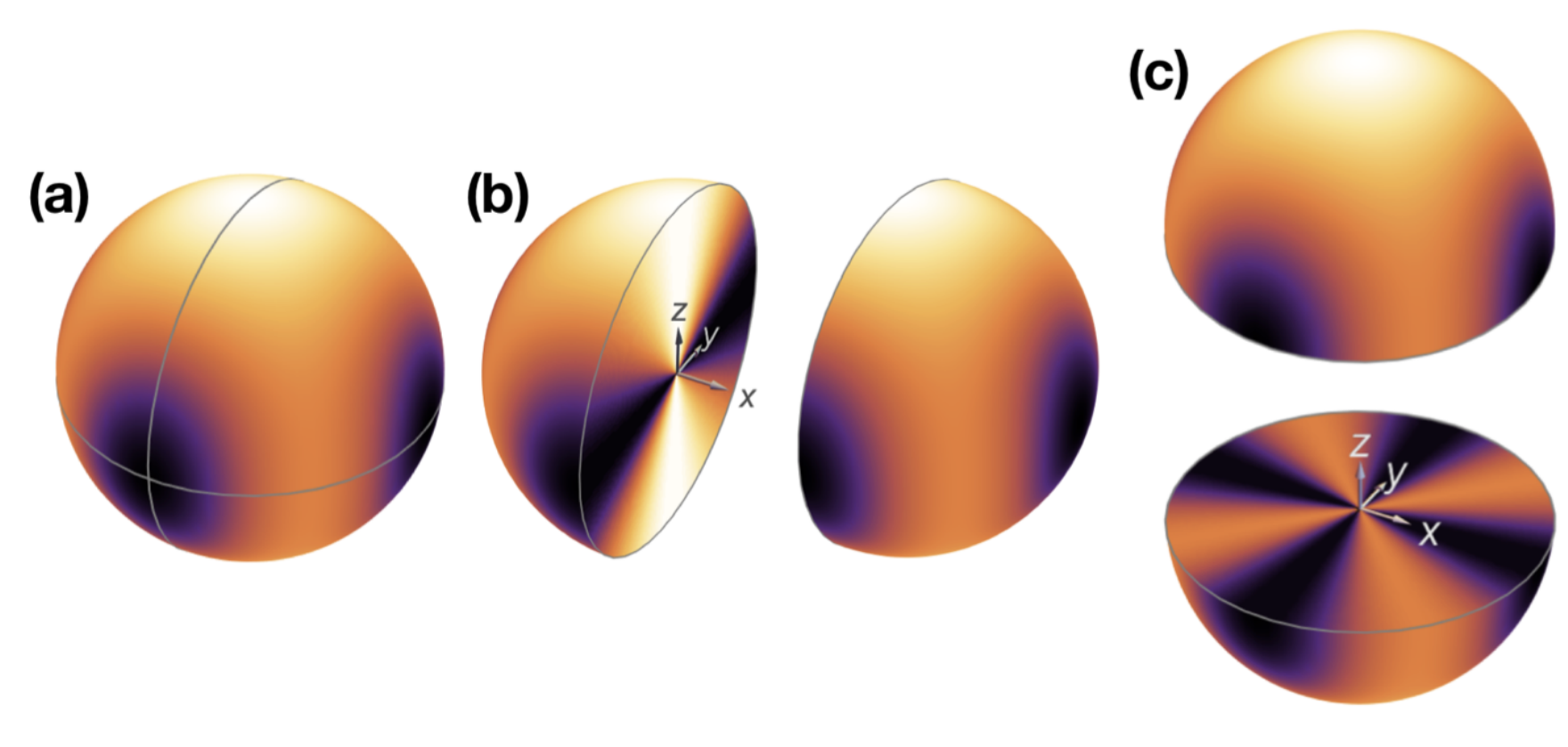}
    \caption{Structure of pinch point singularities from Ref.~\cite{rank2ice} by measuring the correlation function $\langle E_{xy}(q)E_{xy}(-q)\rangle$.}
    \label{fig:pinch}
\end{figure}

Apart from the gapped fracton topological ordered state represented by stabilizer codes, there has also been a parallel search on gapless fracton phases whose low energy effective theory is characterized by tensor gauge theories. Such theories are generalizations of emergent electrodynamics with a close connection to emergent gravity and holography.  In particular, Ref.~\cite{xu} proposed a traceless rank-2 symmetric gauge theory from interacting quantum rotors in 3D with soft graviton excitations.  More recently, motivated by Yb-based materials with a quantum spin ice-like structure, Ref.~\cite{rank2ice} proposed a materially–relevant microscopic model which can potentially realize a traceless rank-2 symmetric gauge theory. Such Yb-based materials can be described by a spin-1/2 Hamiltonian with Heisenberg antiferromagnetic interactions on a breathing pyrochlore lattice with weak Dzyaloshinskii-Moriya (DM) interactions. In addition, such a fracton spin liquid state exhibits 4–fold pinch point singularities in certain spin-spin correlation functions (Figure.~\ref{fig:pinch}) \cite{rank2ice,pinch} which can be verified in polarized neutron scattering experiments.

Finally, we note that there have been numerous other constructions of fracton spin models which we have not had space to discuss here \cite{petrova,brazil,tian2019generalized}.

\section{Foliation}
\label{sec:foliation}

In this section, we describe a set of powerful theoretical tools for characterizing different types of fracton models.  In particular, we focus on the foliation framework, which provides important insights into understanding a variety of fracton phases.  We conclude with a discussion of more recent developments in characterizing fracton systems.

\subsection{Basic Idea}
\label{sec:def_FFO}

\begin{figure}[htbp]
    \centering
    \includegraphics[width=0.5\columnwidth]{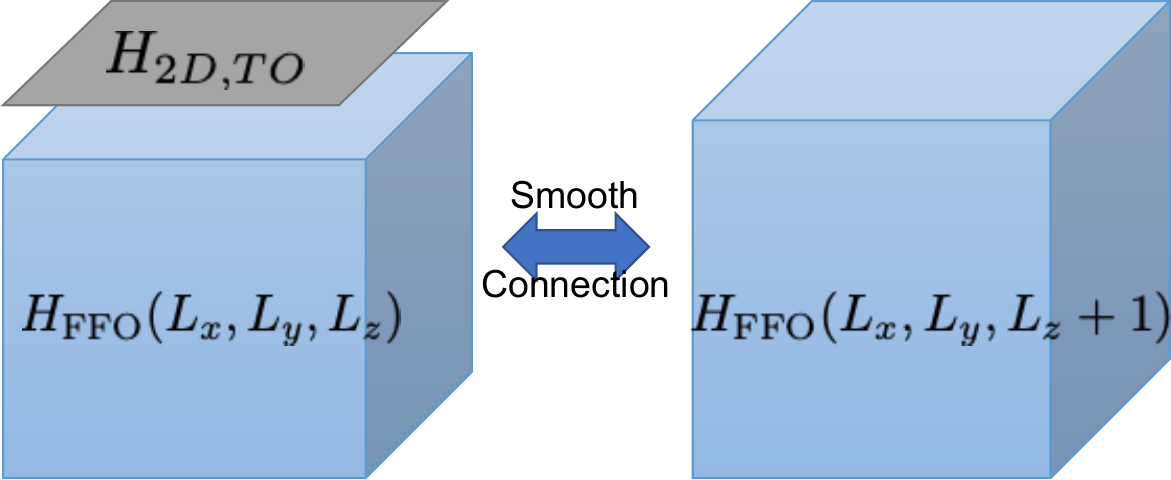}
    \caption{A (type I) fracton model is said to have Foliated Fracton Order (FFO) if models with different system sizes can be smoothly deformed into each other after attaching decoupled 2D layers with topological order (TO). Here the smooth connection can be realized with finite depth quantum circuit on the ground state.}
    \label{fig:FFOa}
\end{figure}

\begin{figure}[htbp]
    \centering
    \includegraphics[width=0.8\columnwidth]{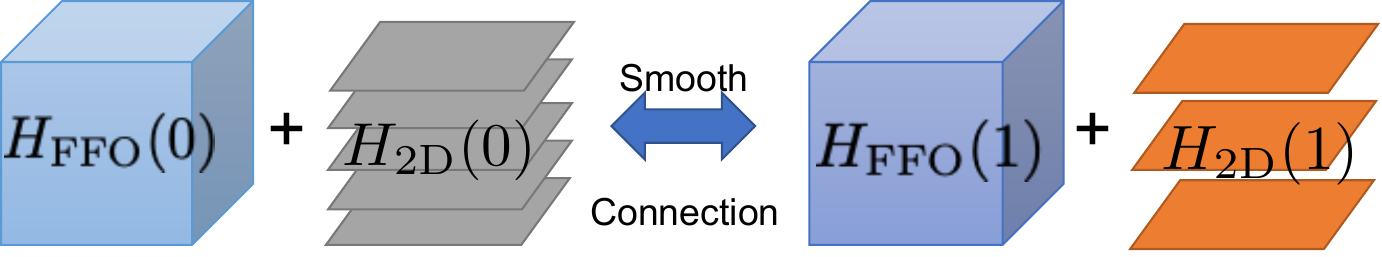}
    \caption{Two foliated fracton models are said to have the same foliated fracton order if they can be smoothly deformed into each other after attaching decoupled 2D layers with topological order. Here the smooth connection can be realized with finite depth quantum circuit on the ground state.}
    \label{fig:FFOb}
\end{figure}

Among the type-I fracton models, it has been shown by Shirley, Slagle, and Chen that many of them have a hidden `foliation' structure and are said to have `foliated fracton order' (FFO) \cite{fol1,fol2}.  That is, starting from a model with a larger system size, we can apply a finite depth local unitary transformation and map the model to a smaller system size together with decoupled layers of 2D gapped states, as illustrated in \figref{fig:FFOa}. As there should be no fundamental change in the order of the system simply due to the change in system size, we should think of the 2D gapped states as free resources in the study of these 3D fracton models even though the 2D gapped states can have highly nontrivial topological order of their own. Correspondingly, we define two foliated fracton models to have the same `foliated fracton order' if they can be related through a finite depth local unitary transformation upon the addition of decoupled stacks of 2D layers of gapped states, as shown in \figref{fig:FFOb}. According to this definition, a stack of 2D topological states has trivial foliated fracton order because it is equivalent to having nothing at all. A nontrivial foliated fracton model has a lot of 2D layers hidden inside of it, yet it is not simply equivalent to a stack.  (See also Ref. \cite{dua2019bifurcating} for a generalized notion of equivalence of fracton phases based on a bifurcating entanglement renormalization scheme.)

This idea of `foliated fracton order' generalizes the notion of topological order, which captures a wide range of nontrivial phenomena in gapped systems in 2D and higher. A model with topological order is `liquid'-like, in the sense that system size can grow by adding decoupled product states and smoothly deforming the model with a finite depth quantum circuit\cite{LULRE}. A model is said to have trivial topological order if it is equivalent to a product state under such deformation and two topological orders are said to be equivalent if they can be deformed into each other after adding product states. 

While definitions are accurate, they are not easy to use. For topological order, a set of universal properties were found, such as ground state degeneracy, topological entanglement entropy, fractional excitation and statistics, which helped to identify and compare topological order in different models. In \secref{sec:properties}, we are going to discuss a similar set of universal properties for FFO and show how they can be used to compare models.

\subsection{General Three-Dimensional Manifolds}

One thing the foliation structure allows us to do is to write down exactly solvable models for certain fracton models on different three dimensional manifolds. This is a natural thing to do because for topological models, a great deal can be learned by putting models on different manifolds and see how their ground state degeneracy changes with the change in topology. In \Ref{fol1}, this was done for the X-cube model, from which we see that FFO models care not only about the topology of the manifold but also about the foliation structure. On the other hand, it is hard to use the ground state degeneracy as a universal characterization of the order, for reasons explained below.

In particular, we construct a lattice by embedding a large number of transversely intersecting surfaces, referred to as leaves, into the 3-manifold $M$. Vertices of the resulting lattice lie at triple intersection points of leaves, while edges lie along the intersections of pairs of leaves; a qubit is placed on each edge. We assume that the location of the leaves are generic enough such that no three leaves intersect along the same line.
The cubic lattice on the 3-torus can be viewed in this way as three orthogonal stacks of toroidal leaves---the $xy$, $yz$, and $xz$ planes.
Unlike the cubic lattice, the general construction may result in some number of non-cubical cells.
Crucially, however, every vertex in this type of lattice is locally isomorphic to a cubic lattice vertex.
This fact allows the X-cube Hamiltonian to be defined as per \eqnref{eq:H}, which we copy below
\begin{equation}
H = -\sum_v \left(A_v^{x}+A_v^{y}+A_v^{z}\right) -\sum_c B_c   
\end{equation}
Similar to the cubic lattice, the three cross operators $A^\mu_v$ are products of $Z$ operators over the four edges emanating from $v$ in the leaf labeled by $\mu$.
The $B_c$ operator is in general a product of $X$ operators over all edges of the 3-cell $c$.
The lattice geometry ensures that the terms in the Hamiltonian are mutually commuting.

The structure of the excitation types and fusion properties carries over from the cubic lattice version of the X-cube model. The notion of dimension-1 and dimension-2 particles is revised in a natural way. In the general lattice construction, dimension-1 particles created at the ends of open string operators are freely mobile along the intersection lines of pairs of surfaces. Furthermore, dimension-2 particles, such as fracton dipoles, are free to move along leaves that are orthogonal to the direction of the dipole moment.

The ground state degeneracy of the model was found to depend not only on the topology of the manifold, but also on the foliation structure (their number, topology, etc). For example, a spherical leaf does not contribute to ground state degeneracy while a torus leaf contributes an additive part of $2$ to the logarithm of ground state degeneracy. However, the ground state degeneracy may not be stable against local perturbations because, unlike the cubic lattice in 3-torus where all non-contractible loops have infinite size, in other manifolds or with other foliations it may happen that non-contractible loops have finite length. Under perturbation, degeneracy coming from such loops will be lifted. Because of this, ground state degeneracy cannot be used as a good quantum number to describe foliated fracton order.

\subsection{Universal Properties and Relation Between Models}
\label{sec:properties}

The definition of FFO given in \secref{sec:def_FFO} applies not only to exactly solvable models, but to generic non-exactly solvable models as well. Based on this definition, one can find universal properties, including entanglement measures and properties of fractional excitations, of the foliated fracton models that remain invariant under both finite depth quantum circuit and the addition of 2D gapped layers. Because there are 2D layers intrinsically hidden in the foliated fracton models, this also means that we need to define universal properties in a way that mods out the contribution coming from 2D layers. 

\begin{figure}[htbp]
    \centering
    \includegraphics[width=0.6\columnwidth]{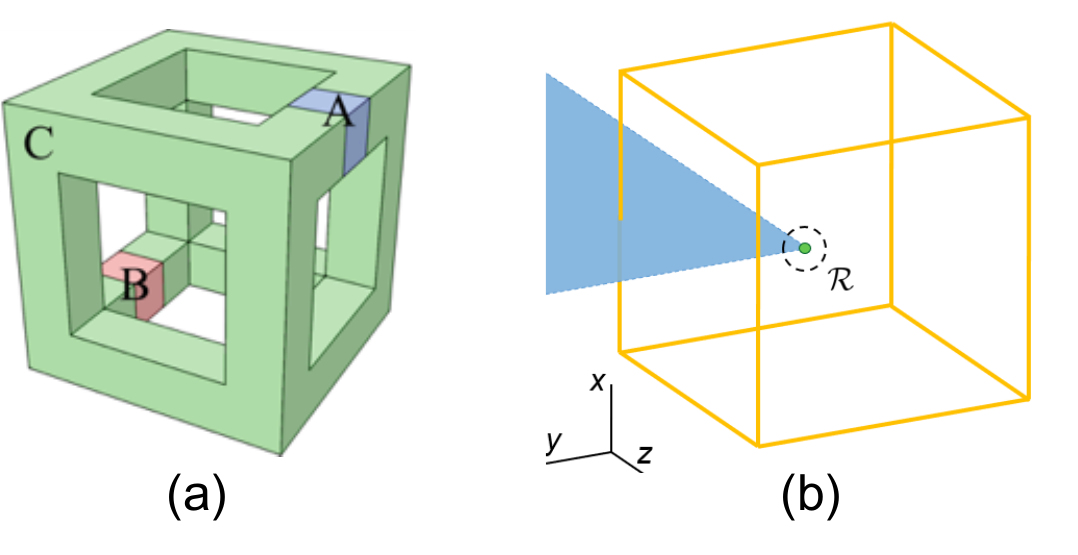}
    \caption{Universal properties of foliated fracton order: (a) Wire-frame structure used for calculating entanglement measure as in \eqnref{eq:EE}; (b) Fractional statistics is obtained by applying interferometric operators around a local excitation in region $R$ (shape of the interferometric operator may differ from that shown in the figure).}
    \label{fig:properties}
\end{figure}

For example, \Ref{fol2} proposed an entanglement measure which cancels both the area law and the sub-leading linear part of the entanglement entropy of a sub-region in a gapped 3D model and retains a constant term. The area law term is generic in a gapped system and depends on details of the model, hence non-universal. The sub-leading area law term is indicative of the foliation structure hidden in the model -- if we have a stack of decouple 2D topological layers, entanglement entropy of a sub-region would have a sub-leading linear term coming from the constant topological entanglement entropy\cite{S_topWen, S_topKitaev} of each layer. The constant term after canceling both is hence a characterization of the underlying FFO. To achieve this, we take a `wire-frame' sub-region whose shape is determined by the foliation structure of the model. For example, for the X-cube mode with foliation layers in $xy$, $yz$, and $zx$ directions, a cubic wire-frame is used with $A$, $B$, $C$ sub-regions (\sfigref{fig:properties}{a}). The entanglement measure is
\begin{equation}
    S_{FFO} = S_A + S_B + S_C - S_{AB}-S_{BC}-S_{AC}+S_{ABC}
    \label{eq:EE}
\end{equation}
This can be applied to various fracton models for comparison. For the X-cube model, $S_{FFO}=1$. 

Besides entanglement entropy, we can also look at fractional excitations. Fractional excitations in topological systems were sorted according to super-selection sectors -- two excitations are considered equivalent if they can change into each other by adding / removing non-fractional local excitations. But this is too coarse for FFO models, as FFO models host an infinite number of super-selection sectors. Instead, we define the notion of `quotient super-selection setors' by modding out not only non-fractional excitations but also dimension-2 fractional excitations which come from foliation layers. The number of sectors is then greatly reduced. For example the X-cube model has one fracton sector and one lineon sector in each direction. We further discussed fractional statistics which are invariant to the addition of 2D layers by applying specially designed interferometric operators around a local excitation (\sfigref{fig:properties}{b}) so that the resulting phase factor is independent of the attachment of dimension-2 particles to the excitation. 

With these universal properties defined, we can now compare different models and see if they may potentially have the same foliated fracton order. In particular, we find that the X-cube model, the semionic X-cube model\cite{hanlayer} and the Majorana checkerboard model\cite{vhf1} all have the same foliated order while the checkerboard model\cite{vhf2} is equivalent to two copies of the X-cube model. The equivalence in the above universal properties is a necessary condition for equivalence of FFO but may not be sufficient. To rigorously establish the equivalence, we found the 2D layers that need to be inserted and explicit local unitary transformations to map one model to another. 

\subsection{Twisted Phases}

With many of the known type I models found to be in the same FFO phase as the X-cube, it is natural to ask whether there exists models with a different FFO order. The answer is yes, as shown in \Ref{fol6} where `twisted' foliated fracton models were constructed. The models are called `twisted' in the same sense that the 2D double-semion model is called a twisted $Z_2$ gauge theory while the 2D Toric code is an un-twisted $Z_2$ gauge theory. The X-cube model can be interpreted as the gauge theory of trivial paramagnet with subsystem planar symmetries\cite{vhf2}, while the twisted models are gauge theories of non-trivial paramagnets with subsystem planar symmetries.  Similarly, Haah's code is closely associated with a gauged fractal symmetry \cite{ungauging}.

One of the twisted FFO models is constructed using the coupled layer construction discussed in \secref{sec:CL} with layers of 2D twisted $Z_2\times Z_2$ gauge theory models. The coupling binds the corresponding $Z_2$ flux strings from intersecting plane together; the resulting model is similar to the X-cube model in that it is has foliation layers in $xy$, $yz$, $zx$ directions and there are fracton, lineon (dimension-1), and dimension-2 particles. On the other hand, it was shown to host a different foliated fracton order.

The twisted model actually behaves in the same way in terms of the entanglement measure and quotient super-selection sector defined in \secref{sec:properties} (indicating that these indeed do not form a complete list of universal properties). Their difference shows up in the dimension-2 particles. To see the difference, we compactify the model from 3D to 2D (by making the $z$ direction finite) so that only dimension-2 particles in the $xy$ plane remain as fractional excitations. By studying their fusion and braiding statistics and compare with what we get from the X-cube model with the same process, we can show that the two sets of dimension-2 particles cannot be mapped into each other by adding 2D layers and local unitary transformation, hence establishing the difference in FFO for the original non-compactified models.  See also Ref. \cite{compactify} for a discussion of calculating invariants for fracton phases based on compactification.

\subsection{New Approaches to Characterizing Fracton Systems}

As we have established, foliation is a powerful tool for characterizing fracton phases, providing many important insights.  However, the notion that two fracton phases are equivalent up to the addition of two-dimensional topological phases is slightly more coarse-grained than the traditional notion of phases of matter, allowing for the possibility that two distinct fracton phases may look identical within the foliation framework.  It is therefore useful to consider other characterizations which capture the more detailed distinctions between fracton phases.  One promising idea along these lines is to characterize fracton phases in terms of their quasiparticle context, along with their associated fusion theory and statistical processes, in direct analogue to the data characterizing more conventional topological phases.

In Reference \cite{fusion}, Pai and Hermele constructed a fusion theory capable of describing the quasiparticle content of fracton phases, along with various examples of nontrivial statistical processes.  The key idea in this fusion theory is to consider the action of translation on the superselection sectors of the theory, which encodes the mobility of quasiparticles.  However, the number of superselection sectors in a fracton theory is infinite, so one must find some organizing principle for these sectors in order to yield a useful fusion theory.  The necessary structure is provided by the conservation laws of the theory, which are in direct correspondence with superselection sectors and can be regarded as an additive group.  For example, the X-cube model is characterized by conservation of charge (mod 2) on every plane normal to a cardinal direction.  There is then an injective mapping $\pi: S\rightarrow P$ from the group of superselection sectors $S$ into the group of plane charges $P$.  (Note, however, that the mapping is not surjective since not all combinations of plane charges can be consistently realized.)  Similar considerations hold for other types of fracton theories.

Armed with this description of $S$ in terms of the conservation laws, we now consider the action of translations on the superselection sectors.  Specifically, we consider the action of a discrete lattice translation, $t_a \in T\cong \mathbb{Z}^3$, where $t_a$ is a translation by lattice vector $a$.  We can construct a mapping $T\times S \rightarrow S$ which serves as a group action of $T$ on $S$, satisfying various physical assumptions.  For example, we can stipulate that $t_a(s_1+s_2) = t_as_1 + t_as_2$, reflecting the fact that it does not matter whether we fuse two particles then translate them, or translate them first and then fuse them.  Furthermore, there is a natural action of $\mathbb{Z}[T]$ (the group ring of translations with integer coefficients) on $S$, which makes $S$ into a $\mathbb{Z}[T]$-module.  (Integer multiplication can be defined via $2t_a s = t_a s + t_a s$, and so on.)  This formulation neatly encodes the mobility of quasiparticles as follows.  For a given $s$, one can identify the subgroup of translations $T_s \subset T$ for which $T_s s = s$.  If translations in a particular direction leave a particle's superselection sector invariant, then the particle is mobile in that direction.  A particle with $T_s \cong T$ is fully mobile, while a particle with trivial $T_s$ is an immobile fracton.  Similarly, $T_s \cong \mathbb{Z}$ and $T_s \cong \mathbb{Z}^2$ indicate lineons and planons, respectively.  It can readily be checked that this formalism correctly captures the immobility of fractons and two-dimensional nature of dipoles in the X-cube model \cite{fusion}.  Furthermore, this logic can even be extended to construct fusion theories of gapless fracton models, such as the $U(1)$ gauge theories.

\begin{figure}
    \centering
    \includegraphics[width=.45\columnwidth]{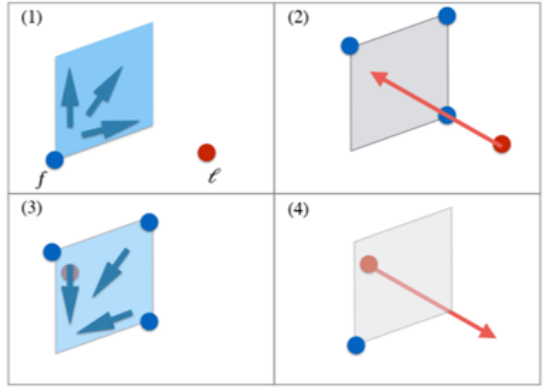}
    \caption{The X-cube model possesses a nontrivial statistical process of fractons and lineons, in which a lineon pierces a fracton membrane operator.  (Figure taken from Ref \cite{fusion}.)}
    \label{fig:braid}
\end{figure}

In addition to fusion, this framework can be used to study various statistical processes in fracton theories, which in general cannot be associated with braiding between two quasiparticle types, but rather must be thought of in terms of a more general sequence of local moves (see also Ref. \cite{bndbraid} for a discussion of statistical processes in fracton systems with a boundary, and Ref. \cite{sorting} for an approach based on a generalized $S$-matrix.).  For example, in the X-cube model, there is a nontrivial statistical phase factor associated with the process shown in Figure \ref{fig:braid}, involving a lineon and multiple fractons.  An analysis of similar statistical processes can be used to demonstrate that the standard X-cube model and its semionic variant represent two separate phases, a distinction which is not captured by the foliation framework.  However, it remains an open question how one can attach a complete set of statistical data to a given fusion theory to fully characterize a fracton model.

\section{Realization in Elasticity Theory}
\label{sec:elasticity}

\subsection{Fracton-Elasticity Duality}

\begin{figure}
    \centering
    \includegraphics[width=.5\columnwidth]{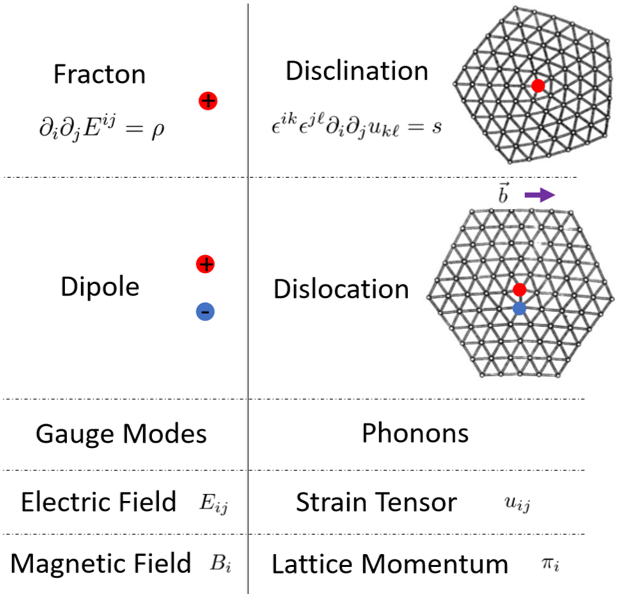}
    \caption{Summary of the duality between elasticity of crystals and a fracton tensor gauge theory.  Fractons and dipoles map to disclinations and dislocations, respectively, while the gapless gauge modes map onto acoustic phonons.  (Figure taken from Reference \cite{elasticity}.)}
    \label{fig:dict}
\end{figure}

While the models we have considered so far have taken the form of complicated spin models, without an immediate connection to material realization, it is important to note that fractons have a much more down-to-earth physical realization as the topological lattice defects of ordinary two-dimensional solids.  This connection between fractons and lattice defects can be seen by studying the conventional elasticity theory of two-dimensional crystals, which turns out to have an exact duality mapping with the scalar charge fracton tensor gauge theory (enriched by an extra global symmetry) \cite{elasticity,elasticityprb}.  Within this duality, disclination defects play the role of immobile fractons, while dislocation defects act as dipoles exhibiting one-dimensional motion.  Meanwhile, the phonons of the crystal map onto the gapless gauge modes of the symmetric tensor gauge theory.\footnote{Importantly, the resulting tensor gauge theory is noncompact, which allows it to maintain stability even in two spatial dimensions.}  The details of this duality are summarized in Figure \ref{fig:dict}.

This duality mapping can be derived through a few simple algebraic manipulations, proceeding as a natural tensor analogue of the more-familiar particle-vortex duality.  The starting point is the usual elastic description of a crystal in terms of a displacement vector field $u_i(x)$, characterizing the displacement of atoms away from their equilibrium positions \cite{landau,chaikin,kardar}.  To lowest order in derivatives, the most general low-energy effective action for a crystal can be written as:
\begin{equation}
S = \int d^2xdt\frac{1}{2}\bigg( (\partial_tu_i)^2-C^{ijk\ell}u_{ij}u_{k\ell}\bigg)
\end{equation}
where $u_{ij} = \frac{1}{2}(\partial_iu_j + \partial_ju_i)$ is the symmetric strain tensor.  Note that the antisymmetric strain $\epsilon^{ij}\partial_iu_j$, representing the local rotation of the crystal, does not appear to lowest order in the action.  This is a consequence of the underlying spontaneously broken rotational symmetry, which dictates that there is no energy cost associated with rotating the crystal as a whole.  (This can also be seen more explicitly in an alternative formulation of elasticity theory \cite{elasticityprb,mikeleo}.)  This action describes the behavior of two gapless modes, corresponding to transverse and longitudinal phonons.  Additionally, a crystal hosts disclination defects which serve as a source for the symmetric strain tensor via:
\begin{equation}
\epsilon^{i\ell}\epsilon^{jk}\partial_\ell\partial_k u_{ij} = \rho
\label{discdef}
\end{equation}
where $\rho$ is the disclination density.  Besides these fundamental topological defects, a crystal also hosts stable dipolar bound states of disclinations, which correspond to dislocation defects.  The stability of these dipolar states is an important clue in making the connection with fracton physics.

The mapping onto fracton physics can be accomplished through what is essentially a simple Hubbard-Stratonovich transformation \cite{elasticity}.  To this end, we introduce the variables $\pi_i$ and $\sigma_{ij}$, corresponding to the lattice momentum and stress tensor, in terms of which we write the action as:
\begin{equation}
S = \int d^2xdt\bigg[\frac{1}{2}C_{ijk\ell}^{-1}\sigma^{ij}\sigma^{k\ell}-\frac{1}{2}\pi^i\pi_i -\sigma^{ij}(\partial_i\Tilde{u}_j + u_{ij}^{(s)}) + \pi^i \partial_t(\Tilde{u}_i+u_i^{(s)}) \bigg]
\end{equation}
where $\Tilde{u}_i$ is the smooth piece of $u_i$, while $u_i^{(s)}$ is the piece corresponding to topological defects.  This action is now linear in the smooth piece $\Tilde{u}_i$.  Integrating out $\Tilde{u}_i$ and changing variables to $B^i = \epsilon^{ij}\pi_j$ and $E^{ij} = \epsilon^{ik}\epsilon^{j\ell}\sigma_{k\ell}$, some straightforward algebra \cite{elasticity} yields an action of the form:
\begin{equation}
S = \int d^2xdt\bigg[\frac{1}{2}\tilde{C}_{ijk\ell}^{-1}E^{ij}E^{k\ell} - \frac{1}{2}B^iB_i - \rho\phi - J^{ij}A_{ij}\bigg]
\label{dual}
\end{equation}
where we have taken advantage of the fact that Newton's equation of motion, $\partial_t\pi^i -\partial_j\sigma^{ij} = 0$, maps onto a tensor Faraday's equation, $\partial_tB^i + \epsilon_{jk}\partial^jE^{ki} = 0$, which allows us to write the fields $E^{ij}$ and $B^i$ in a potential formulation as:
\begin{equation}
E^{ij} = -\partial_tA^{ij} -\partial^i\partial^j\phi\,\,\,\,\,\,\,\,\,\,\,\,B^i = \epsilon_{jk}\partial^jA^{ki}
\end{equation}
We now see that the dual gauge formulation of elasticity theory is precisely the scalar charge tensor gauge theory discussed in detail in Section \ref{sec:tensor}.  In particular, Equation \ref{discdef} defining the disclination density maps onto a Gauss's law given by $\partial_i\partial_jE^{ij} = \rho$, which implies that disclinations exhibit restricted mobility via the conservation of dipole moment.  We can therefore immediately conclude that the disclination defects of two-dimensional crystals behave as fractons.

At this point, we must resolve one important remaining issue with this duality.  The scalar charge tensor gauge theory of Equation \ref{dual} has immobile fracton excitations, which coincides with the fact that disclination motion serves as a source for dislocations.  However, the tensor gauge theory also hosts dipole excitations which at first glance appear to be fully mobile, in contrast to the fact that dislocations only move along their Burgers vector.  To address this tension, we must take into account some additional microscopic information about crystals.  Specifically, crystals are made up of atoms, and the number of atoms is exactly conserved in all processes, corresponding to an underlying $U(1)$ symmetry.  Furthermore, dislocation ``climb" ($i.e.$ motion along the forbidden direction) can occur through emission of vacancy or interstitial defects, corresponding to misplaced atoms in the crystal.  If atom number were not conserved, then a dislocation would be able to move freely in all directions.  This indicates that the mobility restrictions on dislocations are enforced by the presence of an extra global $U(1)$ symmetry in a manifestation of ``symmetry-protected" fracton behavior \cite{supersolid,potter}.  This global symmetry remains present in the dual gauge theory, enforcing one-dimensional behavior on the dipoles.

This duality sheds important light on the phase diagram of the fracton tensor gauge by mapping onto the familiar problem of two-dimensional melting.  As a two-dimensional crystal is heated, it first partially melts into a hexatic phase via the proliferation of dislocation defects, destroying translational order but maintaining rotational order.  As the system is heated further, the hexatic phase eventually melts into an ordinary isotropic liquid via proliferation of disclination defects, destroying the rotational order.  Via the duality mapping, we can then conclude that this fracton tensor gauge theory will exhibit two thermal phase transitions as the temperature is raised, corresponding to the proliferation of dipoles followed by the proliferation of fractons.  In turn, the duality allows the fracton formalism to shed additional light on the phase diagram of two-dimensional crystals.  For example, the duality has been used to provide a simplified derivation of the Halperin-Nelson-Young theory of two-dimensional melting \cite{z3}.  It has also been proposed that the duality may aid in the classification of interacting topological crystalline insulators \cite{elasticity,elasticityprb}.

\subsection{Extensions}

While the original fracton-elasticity duality applies to simple two-dimensional crystals, there are various extensions of this duality to other types of crystals, often with interesting implications for fracton physics.  We here describe some of the most prominent recent developments in this area.

\subsubsection{Three-Dimensional Crystals and Fractonic Lines}

Since the topological defects of two-dimensional crystals behave as fractons, it is natural to ask whether similar physics holds in three dimensions.  However, one quickly encounters the complication that the topological defects of three-dimensional crystals are not point-like, but rather take the form of line-like objects.  Nevertheless, it has been shown that these line-like objects exhibit the restricted mobility of fractons, and are therefore referred to as ``fractonic lines" \cite{3delasticity}.  These mobility restrictions are well-captured by a tensor gauge dual, which in this case is written in terms of a rank-4 tensor gauge field $A_{ijk\ell}$ which is symmetric under $(ij)\leftrightarrow (k\ell)$ and antisymmetric under $i\leftrightarrow j$ and $k\leftrightarrow \ell$.  In other words, this theory combines the properties of symmetric tensor gauge theories (describing point-like fractons) with those of higher form gauge theories (describing extended objects).  A gauge dual effective action for three-dimensional elasticity theory can then be written as:
\begin{equation}
S=\int d^3xdt\bigg[\frac{1}{2}\Tilde{C}^{-1}_{ijk\ell pqrs}E^{ijk\ell}E^{pqrs} - \frac{1}{2}B^{ij}B_{ij} - \rho^{ij}\phi_{ij} - J^{ijk\ell}A_{ijk\ell}\bigg]
\end{equation}
where $E_{ijk\ell}$ is the conjugate electric field variable, $B_{ij}$ is a gauge-invariant magnetic field, $\rho_{ij}$ is the charge density of the line-like defects, and $\phi_{ij}$ and $J_{ij}$ are its potential energy and current, respectively.  The charge density $\rho_{ij}$ can be defined in terms of a Gauss's law of the form:
\begin{equation}
\partial_i\partial_kE^{ijk\ell} = \rho^{j\ell}
\end{equation}
which implies that the charge density obeys $\partial_i\rho^{ij} = 0$.  This Gauss's law can then be used to derive various higher moment conservation laws of the theory.  Importantly, however, these are not the usual sort of conservation laws integrated over a three-dimensional region of space.  Rather, these are higher moment conservation laws on the \emph{flux} of $\rho^{ij}$ through two-dimensional surfaces.  In this sense, fractonic lines are governed by the natural higher moment analogues of higher form symmetries \cite{higher1,higher2}.  Since their introduction, various generalizations of these extended fractonic objects have been proposed \cite{extend,premwilliamson,shenoy}.

\subsubsection{Supersolids}

As noted earlier, the one-dimensional behavior of dislocations in a crystal is closely tied to the conservation of particle number.  It then becomes an interesting question to think about the interplay of crystalline order, with its associated fracton behavior, and superfluid order.  Indeed, since superfluids exhibit effective non-conservation of particle number (via spontaneously broken $U(1)$ symmetry), a system featuring coexisting crystalline and superfluid order ($i.e.$ a supersolid) will host fully mobile dislocations \cite{supersolid,potter}.  Fracton-elasticity duality therefore indicates the presence of two distinct fracton phases at zero temperature, corresponding to the solid and supersolid phases, which are distinguished by the mobility of their dipole excitations.  Furthermore, the duality can be extended to include both the crystalline and superfluid sectors, thereby combining fracton-elasticity and particle-vortex dualities into one master bosonic duality.  The resulting gauge dual of a two-dimensional supersolid is written in terms of both a symmetric tensor gauge field $A_{ij}$ and a vector gauge field $a_i$, with an action given by \cite{supersolid}:
\begin{equation}
S = \int d^2xdt\bigg[\frac{1}{2}(\hat{C}_{ijk\ell}E^{ij}E^{k\ell} - \overline{\rho}^{-1}B^iB_i + \overline{K}^{-1}e^ie_i - \overline{\chi}^{-1}b^2) - gB^ie_i -g'E^i_{\,\,i}b + \cdot\cdot\cdot\bigg]
\end{equation}
where the ``$\cdot\cdot\cdot$" represents all source terms for the gauge fields.  The first two terms represent the crystalline sector, the second two terms represent the superfluid sector, and the last two terms represent coupling between the two types of order.  This coupling leads to a subtle interplay between crystalline defects and superfluid vortices, with important consequences for the zero-temperature phase diagram of bosons.  For example, any quantum melting transition of a solid will necessarily induce superfluid order in the resulting liquid phase, whether or not superfluidity is present in the original solid.

\section{Non-Ergodic Behavior in Fracton Systems}
\label{sec:nonergodic}

Now that we have firmly established several physical realizations of fractons, such as excitations of spin models and topological crystalline defects, we now move on to discuss some of the phenomenology of fracton systems.  Perhaps most notably, the limited mobility of fractons places severe restrictions on the ability of a fracton system to reach thermal equilibrium.  Fracton systems generically exhibit slow, glassy dynamics, such that the time to reach thermal equilibrium can become arbitrarily long at low temperatures, in a manifestation of ``asymptotic localization" \cite{glassy}.  In certain special cases, particularly in one dimension, fractons can even exhibit truly non-ergodic behavior, failing to ever reach thermal equilibrium \cite{localization}.  We discuss each of these two situations in turn.

\subsection{Glassy Dynamics of Fractons}

Since a fracton cannot move in isolation, it is tempting to think that it is a trivially localized excitation.  At zero temperature, this is indeed the case, and a single fracton will remain localized at its initial location for infinitely long times.  At finite temperature, however, the story becomes more complicated.  We first focus on type-I fracton models, such as the X-cube model or scalar charge theory, in which fractons can form stable mobile bound states.  While such a fracton cannot move by itself, it can move through the absorption of an additional composite excitation, which we take to be a dipole for concreteness.  At finite temperature, there will be a thermally excited bath of dipoles throughout the system, and a fracton can move by absorbing dipoles from this bath.  As studied by Prem et al., a series of such processes will generically allow a fracton to diffusively delocalize over the entire system, thereby losing the memory of its initial conditions \cite{glassy}.  (See also some important precursor work in References \cite{chamon,kim}.)

While such delocalization processes will generically occur, it is important to note that they are limited by the number of thermally excited dipoles available for absorption by the fracton.  Assuming that there is an energy gap $\Delta$ to create dipoles, and the thermal bath of dipoles is at temperature $T$, then the density of dipoles available for absorption will scale as $\exp(-\Delta/T)$, which in turn sets the scale for the diffusion of fractons.  It then follows that the equilibration time for this system ($i.e.$ the time necessary for an initially localized fracton to disperse around the system) scales as $\exp(\Delta/T)$.  While the system does eventually reach equilibrium, the timescale for thermalization grows exponentially as the temperature is lowered.  At the lowest temperatures, this timescale can be arbitrarily long ($e.g.$ longer than the age of the universe).  Thus, in the low-temperature regime, such a system will be effectively localized for all intents and purposes.  Such a scenario is a manifestation of glassy dynamics, or ``asymptotic localization" \cite{asymptotic}.  For the $U(1)$ theories, featuring long-range interactions between fractons, this slow dynamics also manifests in a delayed onset of screening \cite{u1finite}.  Unlike ordinary charges, which immediately have their long-range fields screened at finite-temperature, fractons can exhibit long-range interactions for extraordinarily long times prior to being fully screened.

Finally, we note that the situation is more complicated for type-II fracton models, such as Haah's code, in which there are no mobile bound states.  In such models, fractons can still move via the emission or absorption of additional composite excitations.  However, these composites are now also strictly immobile.  As such, it is extremely difficult for the composites to form a thermal bath from which the fracton can absorb excitations.  In Reference \cite{glassy}, a detailed study of thermalization in such models was undertaken, finding that they generically exhibit a much slower subdiffusive delocalization of charge.  For these systems, the time for relaxation to equilibrium scales as $\exp(\Delta^2/T^2)$.  This superexponential behavior of the relaxation time indicates that type-II models remember their initial conditions for even longer than their type-I counterparts.

\subsection{Localization of Fractons in One Dimension}

While dipole absorption can lead to thermalization at the longest timescales in generic three-dimensional fracton models, the behavior of one-dimensional fracton systems can be a bit more complicated, leading in some cases to truly non-ergodic behavior.  Such a failure of a one-dimensional fracton system to fail to thermalize at any timescale was first encountered in the context of random unitary circuit dynamics \cite{localization}.  Random unitary circuits provide, in a certain sense, the most generic form of unitary time evolution, without the presence of additional constraining conservation laws, such as energy conservation.  These models have provided a testing ground for ideas about the growth of quantum entanglement and operator spreading \cite{nahum1,nahum2,keyser,jonay}.  In particular, operators generically exhibit ballistic spreading of their support under Heisenberg time evolution.  It is also possible to implement various conservation laws in random unitary circuits, to study how they affect the behavior of operator spreading.  For example, it has been shown that the presence of charge conservation leads to certain operators having a slow diffusively spreading piece, in addition to a ballistic piece \cite{khemanicircuit,keyser2}.

\begin{figure}
    \centering
    \includegraphics[width=.9\columnwidth]{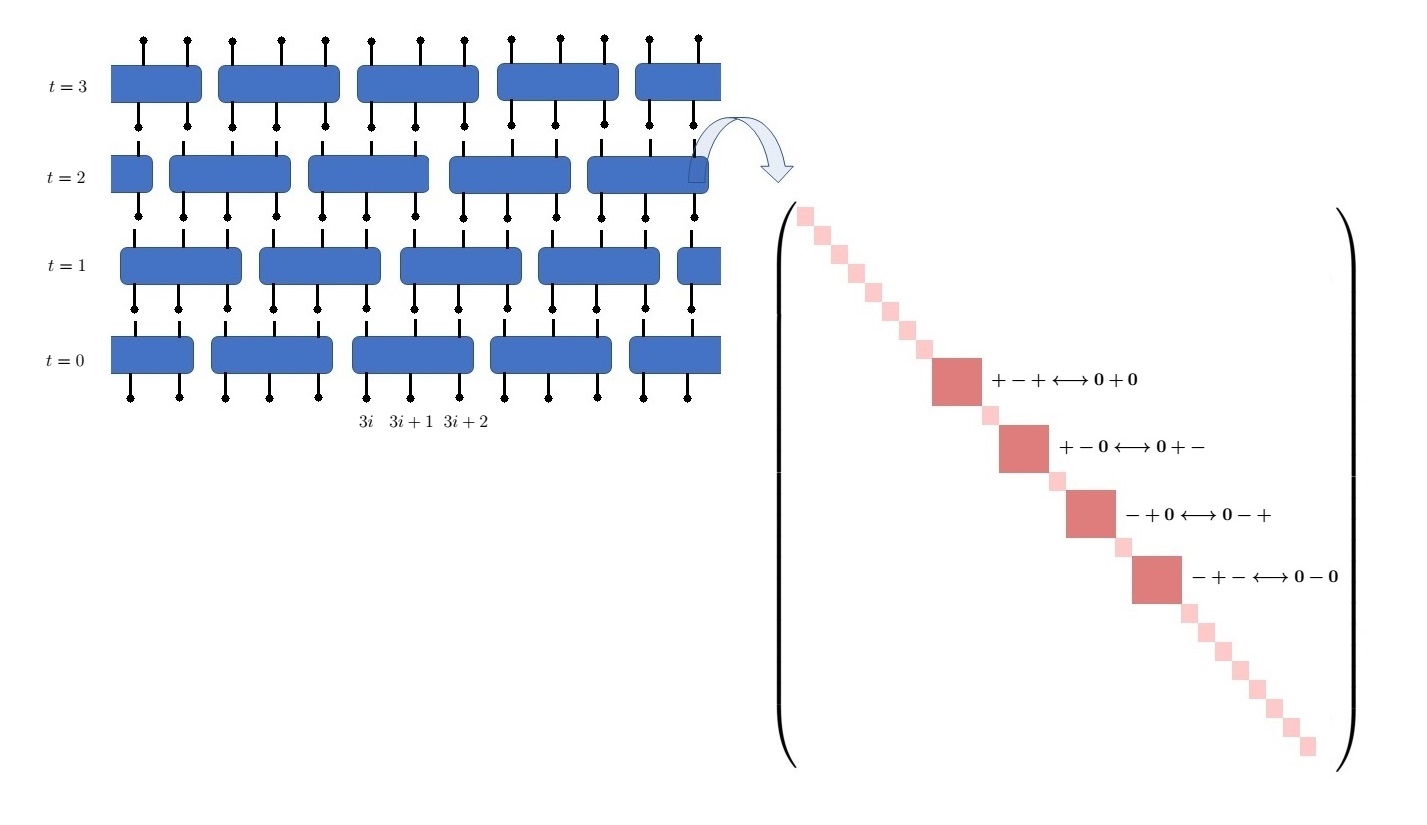}
    \caption{A minimal random unitary circuit obeying fracton conservation laws: Blue blocks represent unitary gates acting on sets of three adjacent spins.  Each gate has a block diagonal structure within sectors of fixed charge and dipole moment.  (Figure taken from Reference \cite{localization}.)}
    \label{fig:rucircuit}
\end{figure}

In Reference \cite{localization}, Pai et al. studied random unitary circuits subject to the two conservation laws of charge and dipole conservation, thereby implementing fracton behavior.  (This model takes fracton behavior as a starting point, in order to study its physical consequences, as opposed to deriving fracton conservation laws from some underyling microscopic interactions.)  The authors constructed a minimal random unitary circuit model consistent with these conservation laws, depicted in Figure \ref{fig:rucircuit}, which consists of layers of local unitary gates acting on a set of spins, where $S_z$ is regarded as the ``charge," as done in \cite{khemanicircuit}.  Notably, in contrast to the circuits studied in previous work, this model requires the presence of three-site gates ($i.e.$ next-nearest-neighbor interactions) to observe any nontrivial dynamics, since there are no nearest neighbor processes consistent with fracton conservation laws.  This circuit was then used to study the Heisenberg time evolution of a fracton number operator initially localized at site $i$:
\begin{equation}
O_i(t) = U^\dagger(t) \bigg(\cdot\cdot\cdot I\otimes I \otimes S^z_i \otimes I \otimes I \cdot\cdot\cdot\bigg) U(t)
\end{equation}
where $U(t)$ is the unitary defined by the circuit.  By looking at the right weight of this operator (a quantity which serves as a measure of its spatial support), it was determined that an $O(1)$ portion of the fracton charge remains localized around its initial position at arbitrarily long times.  Thus, this system exhibits non-ergodic behavior, never forgetting the initial position of the fracton.  Various other aspects of the dynamics were studied in \cite{localization}, such as an anomalous exponent in the ``tail" of the ballistically spreading peak.  It was also shown that fractons could attract each other under random unitary evolution, consistent with the gravitational analysis of Reference \cite{mach}.

Reference \cite{localization} proposed a hydrodynamic explanation for the observed localization, which correctly predicts various aspects of the dynamics, such as the anomalous tail exponent.  However, the generality of these hydrodynamic equations was called into question by later work which showed that the presence of four-site gates ($i.e.$ next-next-nearest neighbor interactions) will eventually cause the fracton operator to almost completely delocalize \cite{pollmann}.  It remains an interesting open question how the coarse-grained hydrodynamic description should be modified to account for this dependence on the range of interactions.  Meanwhile, Reference \cite{pollmann} proposed a more microscopic picture for this localization in terms of ``fragmentation" of the Hilbert space, which implies the presence of ``inert" localized states with trivial time evolution.  (See also parallel work in Ref. \cite{shat}.)  In turn, Reference \cite{scar} showed how these localized states are intricately connected with the many-body scar paradigm, in which a small number of non-thermal states exist in an otherwise thermal spectrum \cite{qmbscars}.  All of this unexpected behavior, arising as a consequence of two simple conservation laws, has led to numerous recent investigations into ergodicity breaking in fracton systems \cite{confinement,krylov,pollmann2,iaconis,taylor,zhicheng}.

Notably, the non-ergodic behavior of fracton systems may prove to be one of the most accessible features in near-term experiments.  It was first noted in Reference \cite{bloch} that the conservation of dipole moment, and therefore fracton behavior, can be effectively enforced via the presence of a strong linear potential.  (In this sense, fracton localization has close ties to older studies on the Wannier-Stark localization of electrons moving in a strong uniform electric field \cite{stark}.)  By implementing linear potentials in ultra-cold atom systems, it should be possible to impose fracton behavior ``by hand," without any corresponding microscopic gauge structure.  For such a system, non-ergodic behavior will serve as the most direct signature of fracton physics.  Indeed, a recent cold-atom experiment has been able to create a linear potential, though not yet of sufficient strength to enforce dipole conservation \cite{huse}.  Nevertheless, this system already exhibits slow subdiffusive transport, which may be a precursor to fracton behavior.  By increasing the strength of the imposed linear potential, it may eventually be possible to experimentally observe the non-ergodic behavior expected for fractons.

\section{Gravitational and Holographic Behavior}
\label{sec:gravity}

One of the more unusual properties exhibited by fractons is the fact that, in certain models, fractons can exert a gravitational force on each other.  This force is ``gravitational" in the sense that it is both universally attractive and encoded in an effective geometry seen by the fractons.  This feature first became clear due to the relationship between fractons and symmetric tensor gauge theories, which also play a central role in general relativity.  Indeed, fracton systems and general relativity share a similar set of conservation laws, leading directly to universal attraction \cite{mach}.  We first go through the basic idea behind these gravitational properties, followed by the more recent development that certain fracton models exhibit holographic properties, leading to toy models for the AdS/CFT correspondence \cite{holo1,holo2,holo3}.

\subsection{Gravitation in Fracton Systems}

Since many fracton models are formulated in the language of rank-2 symmetric tensor gauge theories, it is natural to expect that these systems should have some similarities with general relativity, in which the dynamical metric plays the role of a symmetric tensor gauge field.  In fact, the parallels between fracton theories and gravitational theories run much deeper.  To see this, it is useful to consider the behavior of linearized gravity, in which the metric is expanded around a flat background as $g_{\mu\nu} = \eta_{\mu\nu} + h_{\mu\nu}$, where $\eta_{\mu\nu}$ is the Minkowski metric and $h_{\mu\nu}$ is a small perturbation.  We also choose to work in a gauge which has only the spatial components of the linearized metric, $h_{ij}$.  With these choices, the $00$ component of Einstein's equations can be written as:
\begin{equation}
\partial_i\partial_jh^{ij} - \partial^2 h^i_{\,\,i} = T^{00}
\end{equation}
which takes the form of a Gauss's law, in which the energy density $T^{00}$ acts as a source for the metric.  This equation should be compared directly with the Gauss's law of the scalar charge tensor gauge theory, which led directly to the conservation of dipole moment:
\begin{equation}
\partial_i\partial_j E^{ij} = \rho \,\,\,\,\,\,\,\Rightarrow\,\,\,\,\,\,\, \int d^dx\,(\rho\vec{x}) = \textrm{constant}
\end{equation}
Since general relativity also possesses a double-divergence Gauss's law, we expect that it too should exhibit some form of dipolar conservation law on its energy density.  Indeed, such a conservation law is found in general relativity in the form of conservation of center of mass motion:
\begin{equation}
\int d^dx\,(T^{00}x^i - T^{0i}x^0) = \textrm{constant}
\end{equation}
where $x^0$ is the time coordinate.  This equation, representing the conservation law associated with boost symmetry, reflects the fact that the center of mass of a gravitational system must move at a constant speed.  If we choose a reference frame appropriately, we can eliminate the second term (representing overall motion of the system), leaving us with the conservation of center of mass:
\begin{equation}
\int d^dx\,(T^{00}\vec{x}) = \textrm{constant}
\end{equation}
This conservation law appears essentially identical to the dipolar conservation law governing the scalar charge theory.  We must then resolve the following puzzle:  Why does such a conservation law seem to make fractons immobile, while still permitting gravitational particles to move and attract each other?

\begin{figure}
    \centering
    \includegraphics[width=0.95\columnwidth]{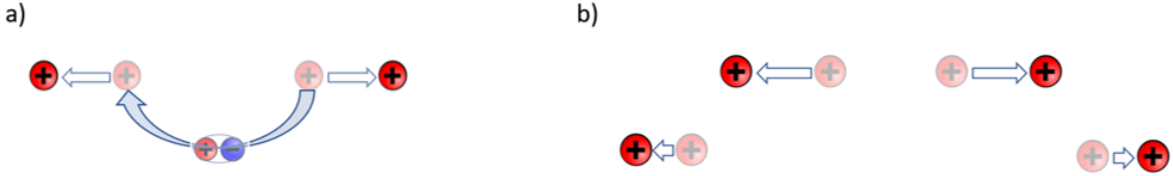}
    \caption{a) Two fractons can push off each other via the exchange of a virtual dipole.  b) Locality of the model dictates that fractons move faster when they are nearby and slower when they are far apart.}
    \label{fig:grav}
\end{figure}

The answer to this riddle lies in the fact that fractons do possess a certain limited amount of mobility in the presence of other fractons.  While it is not possible for a fracton to move in isolation, it is possible for a fracton to move by ``pushing off" a second fracton in such a way that the total dipole moment is conserved, as indicated in Figure \ref{fig:grav}a.  While such a process can allow two fractons to move, the locality of the underlying Hamiltonian dictates that this process must occur via the propagation of some mediating particle between the two fractons.  Specifically, fractons push off of each other via the exchange of a virtual dipole.  The matrix element for this process is then proportional to the propagator of a dipole from the location of one fracton to another.  Importantly, such a propagator decays as a function of the separation between two fractons, such that nearby fractons move more quickly than well-separated ones, as shown in Figure \ref{fig:grav}b.

At this level, it can already schematically be seen that fractons exert an effective attraction on each other, since a pair of fractons will slow down as they begin to move apart.  This logic can be placed on firmer footing by studying the semi-classical equations of motion of the system, which indicate that a fracton moves along geodesic-like curves of an effective geometry dictated by the positions of all other fractons in the system \cite{mach}.  In this sense, the interaction between fractons is purely geometric, just as in general relativity.  Another notable fact about this interaction is that fractons do not strictly speaking exert force on each other, in the sense of exchanging momentum.  Rather, fractons ``exert velocity" on each other via the exchange of position ($i.e.$ center of mass).  In this language, the interaction between fractons is automatically attractive, since locality dictates that well-separated fractons must exert less velocity on each other than nearby fractons.  It is also important to note that generic fracton models, which mimic the behavior of linearized gravity, exhibit only a short-ranged gravitational attraction due to the mass gap associated with dipoles.  Only a fully nonlinear theory of gravity will possess a power-law gravitational attraction.  It is an interesting open question whether a lattice fracton model can be imbued with this property.

Finally, we must resolve one seeming discrepancy between fracton theories and gravitational physics.  While the ``pushing off" mechanism we have discussed allows fractons to move in the vicinity of each other, well-separated fractons effectively become locked in place again, in contrast to the expected behavior of a gravitational theory, in which well-separated particles carry an intrinsic mass.  This difference disappears, however, upon adding a background charge density to fracton models, which endows fractons with a finite mass and promotes them to ordinary gravitational particles.  In this sense, fractons exhibit an explicit example of Mach's principle, which dictates that inertia should be determined by a particle's surroundings rather than being an intrinsic property.  It remains to be seen whether this perspective can shed new light on the structure of more familiar gravitational models.

\subsection{Holography in a Fracton Model}

Since fracton theories can provide toy models for gravity, it should not be surprising that certain fracton models can even serve as toy models for holography \cite{holo1,holo2,holo3}.  In its simplest incarnation, the holographic principle indicates that a gravitational theory defined on anti-de Sitter (AdS) space ($i.e.$ a space with constant negative curvature) has all of its information encoded in a theory defined on its boundary, usually a conformal field theory.  While this duality between the bulk and boundary of a gravitational system was originally encountered in the context of string theory \cite{maldacena}, similar physics has since been observed in much simpler settings.  For example, toy models for holography have been encountered in the context of spin models with quantum error-correcting behavior \cite{happy}.

\begin{figure}
    \centering
    \includegraphics[width=0.3\columnwidth]{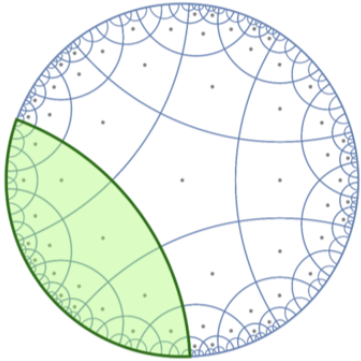}
    \caption{The hyperbolic fracton model permits Rindler reconstruction within the minimal convex wedge (green) from the state along its boundary.}
    \label{fig:wedge}
\end{figure}

Recently, a new type of holographic toy model has been proposed by Yan in the form of a classical fracton model defined on a two-dimensional hyperbolic lattice, mimicking the properties of AdS space \cite{holo1}.  This model, which is closely related to the X-cube model, possesses subsystem symmetries on every geodesic of the lattice, resulting in immobile fracton excitations.  The model also obeys various properties expected of holographic theories.  For example, the mutual information between subregions obeys the expected Ryu-Takayanagi formula.  (Note that, since the model is classical, we must consider mutual information instead of entanglement entropy.)  Furthermore, knowledge of the state on a particular segment of the boundary allows for reconstruction of the bulk state within the minimal convex wedge bounded by that boundary segment, as a manifestation of Rindler reconstruction (see Figure \ref{fig:wedge}).  Via a duality transformation, it can also be shown how this model provides an explicit realization of the bit-thread formulation of holography \cite{holo2,bit}, which in turn hints at a more general perspective on holographic toy models \cite{holo3}.

\section{New Condensed Matter Platforms for Fractons}
\label{sec:cmp}

While a variety of exactly solvable fracton models have been proposed, there is a need for more concrete platforms to realize them experimentally.  In this section, we list several experimental proposals and material fabrications for fracton phases of matter.

\subsection{Majorana Islands}

The theoretical aspects of fracton phases were originally proposed in the context of quantum stabilizer codes and exactly solvable spin models.  However, the direct physical realization of these models remains a key challenge as most stabilizer codes contain complicated spin cluster interactions.  Fortunately, a rich set of quantum spin models can emerge via Majorana quantum Lego whose building blocks are within experimental reach. Here, we briefly mention that many known fracton stabilizer codes can be obtained from such Majorana quantum Lego building blocks.  The principal ingredients of Majorana quantum Lego are Coulomb blockaded Majorana islands and weak inter-island Majorana hybridizations.  Each island contains some number of Majoranas, e.g., at the ends of semiconductor wires proximity coupled to a superconductor \cite{Lutchyn2010,Oreg2010}. The island's charging energy fixes its fermion parity, corresponding to a multi-Majorana interaction.

\begin{figure}[t]
  \centering
    \includegraphics[width=0.7\textwidth]{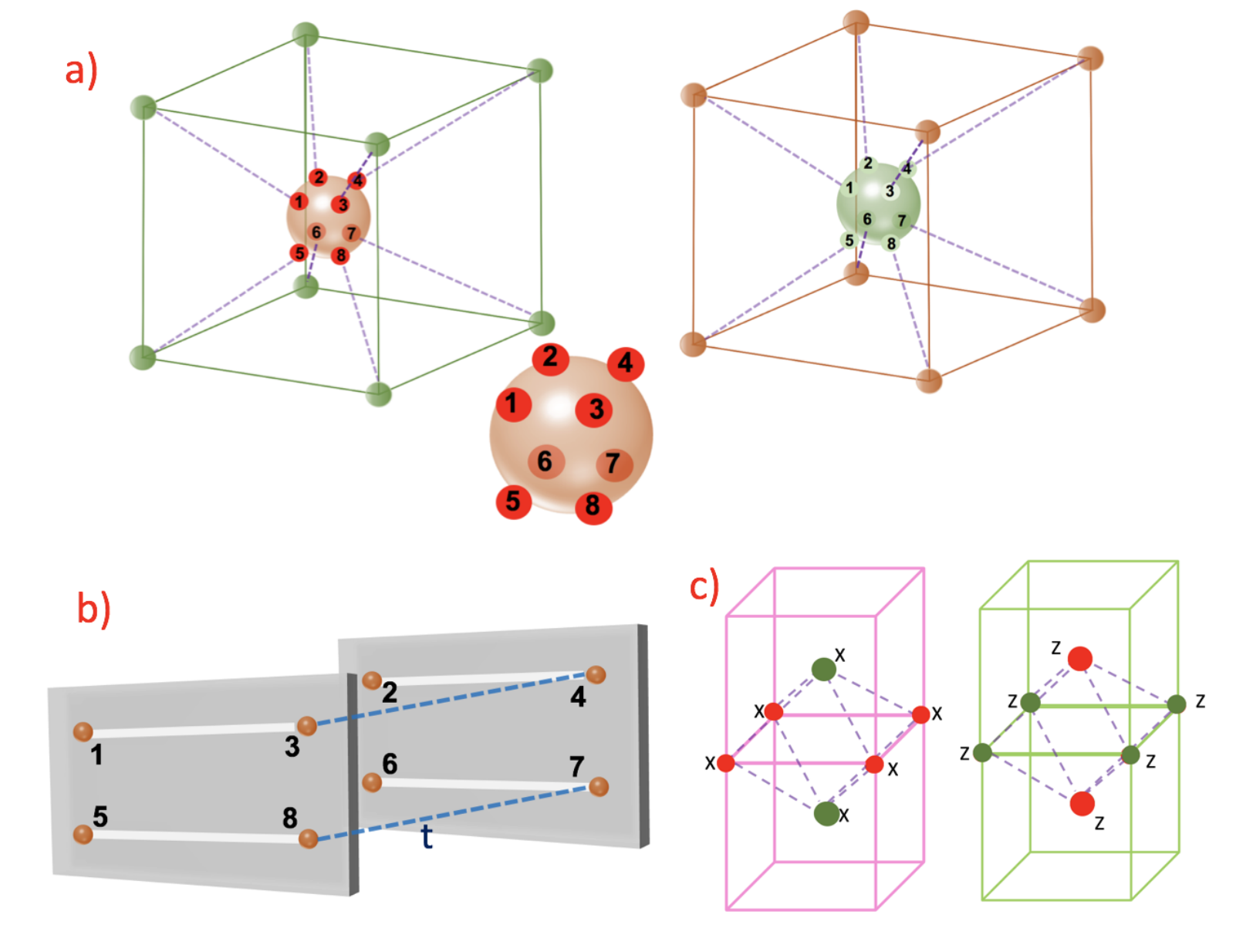} 
  \caption{Construction for the planon-lineon code. a) Body-centered cubic lattice with eight MZM on all corner (green) and center (red) sites. Majorana hybridization is illustrated by dashed lines. b) Setup for realizing the Majorana quartet interaction in Eq.~(\ref{int}). Two Majorana quartets (red dots) are placed on floating superconducting islands, fixing the corresponding fermion parities via charging energy. The third quartet in Eq.~(\ref{int}) is generated by the Majorana hybridizations indicated by the dashed lines. c) The two types of octahedral cells which support the stabilizers of the planon-lineon code.} 
    \label{cx}
\end{figure}

We illustrate an explicit example proposed in Ref.~\cite{majorana} with a topological superconductor on a body-centered cubic lattice. Each site contains eight Majoranas $\gamma^1, \ldots,\gamma^8$ which are each hybridized with a Majorana on a nearest-neighbor site as shown in Fig.~\ref{cx}. Thus, the topological superconductor has the Hamiltonian \begin{align} 
H=-it' \sum_{\langle i,j\rangle} & (\gamma_i^1 \gamma_{j}^7+ \gamma_i^2 \gamma_{j}^8+ \gamma_i^4 \gamma_{j}^5+ \gamma_i^3 \gamma_{j}^6)
\label{H}
\end{align}
and can be thought of as built from crossing one-dimensional Kitaev chains along the $(\pm1,\pm1,1)$ directions. 

We now consider onsite interactions which couple quartets of Majoranas,  
 \begin{align} 
&H_{\rm int}=U(\gamma^1_i \gamma^3_i \gamma^8_i \gamma^5_i+\gamma^3_i \gamma^4_i \gamma^7_i \gamma^8_i+\gamma^4_i \gamma^2_i \gamma^6_i \gamma^7_i).
\label{int}
\end{align}
These interactions suppress hopping of single Majoranas between sites. In the strong-$U$ limit, they project each site into the $\gamma^1_i \gamma^3_i \gamma^8_i \gamma^5_i=\gamma^3_i \gamma^4_i \gamma^7_i \gamma^8_i=\gamma^4_i \gamma^2_i \gamma^6_i \gamma^7_i=-1$ subspace. The product of the three parity constraints also implies $\gamma^2_i \gamma^1_i \gamma^5_i \gamma^6_i=-1$, constraining the Majorana quartets associated with the four vertical faces of the red cube in Fig.~\ref{cx}. 

To implement this interaction in an experimental setup, we distribute the eight Majoranas of each site over two adjacent superconducting islands (SCI) as shown in Fig.~\ref{cx}. Each SCI could be made from two semiconductor quantum wires proximity coupled to the same superconductor. The proximity-coupled quantum wires effectively realize open Kitaev chains with two Majorana zero modes localized at their ends, so that there are a total of four Majoranas on each SCI. By virtue of their charging energy, each SCI can be tuned to have even fermion parity, effectively implementing the interaction terms $U(\gamma^1_i \gamma^3_i \gamma^8_i \gamma^5_i+\gamma^4_i \gamma^2_i \gamma^6_i \gamma^7_i)$ in Eq.~(\ref{int}).  To generate the remaining four-Majorana interaction in Eq.~(\ref{int}), we turn on inter-island Majorana hybridization $H_{t}=it(\gamma^3 \gamma^4+\gamma^8 \gamma^7)$ with amplitude $t$.  These inter-island hybridizations can in principle be implemented by direct tunnel coupling. Alternatively, and perhaps more flexibly, one can bridge between the two Majorana islands using a coherent link. Its two Majorana end states would then be tunnel coupled to the two Majoranas of the Majorana islands which one wants to hybridize. Since the hybridization between the Majoranas on the coherent link and the islands can be realized through gate controlled tunnel junctions, the hybridization strength is tunable. In the limit of a large charging energy, which fixes the fermion parities of the SCI, a single Majorana tunneling between the islands is suppressed and the lowest order processes involve pairs of Majorana tunneling terms as Eq.~(\ref{int}). 

Under such parity constraints driven by interaction, each site retains a single spin-${1}/{2}$ degree of freedom. We can then choose the parities of the top and bottom faces as
the Pauli-$Z$ operator $\sigma_i^z=\gamma^1_i \gamma^2_i \gamma^4_i \gamma^3_i$ and the product of two Majoranas associated with any vertical edge as the Pauli-$X$ operator $\sigma_i^x$, or vice versa. In the strong-$U$ limit, we can treat the Majorana hybridizations as a perturbation. The leading-order Hamiltonian involves 16-Majorana terms for the octahedra shown in Fig.~\ref{cx}. Writing the Hamiltonian in the spin representation yields
\begin{align} 
& H=-\sum_{{\rm octahedra}}\left\{ \prod_{i \in \text{octa}^a} \sigma^x_i + \prod_{i \in \text{octa}^b} \sigma^z_i\right\}.
\label{pl}
\end{align}
Here $\text{octa}^a$ and $\text{octa}^b$ refer to the two types of octahedra in Fig.~\ref{cx} with red (green) sites at top and bottom and four green (red) sites in between. Thus, our construction exactly reproduces the planon-lineon model \cite{fol5} whose elementary quasiparticles are lineons and planons with mobility restricted to the $z$-direction and the $xz$ ($yz$)- planes, respectively. 

This setting \cite{majorana} produces a wide variety of fracton states and promises numerous opportunities for probing and controlling fracton phases experimentally.

\subsection{Plaquette Paramagnets in Two Dimensions}

The wide variety of proposals for fracton models calls for physically-realistic models prone to yield such states. In Ref.~\cite{plaquette}, the authors suggest that some fractonic behaviour can emerge in frustrated quantum paramagnets as a consequence of fluctuating plaquette order or resonating cube order.  Motivated by the precursor work on resonanting valence plaqeutte\cite{rsvp,xu2008resonating} crystal, in this section, we introduce the fractonic properties of topological defects in valence plaquette solid (VPS) phases on square lattices. We show that the defects of the VPS order parameter, in addition to possessing non-trivial quantum numbers, have fracton mobility constraints deep in the VPS phase, which has been overlooked previously. 

\begin{figure}[h]
  \centering
      \includegraphics[width=0.55\textwidth]{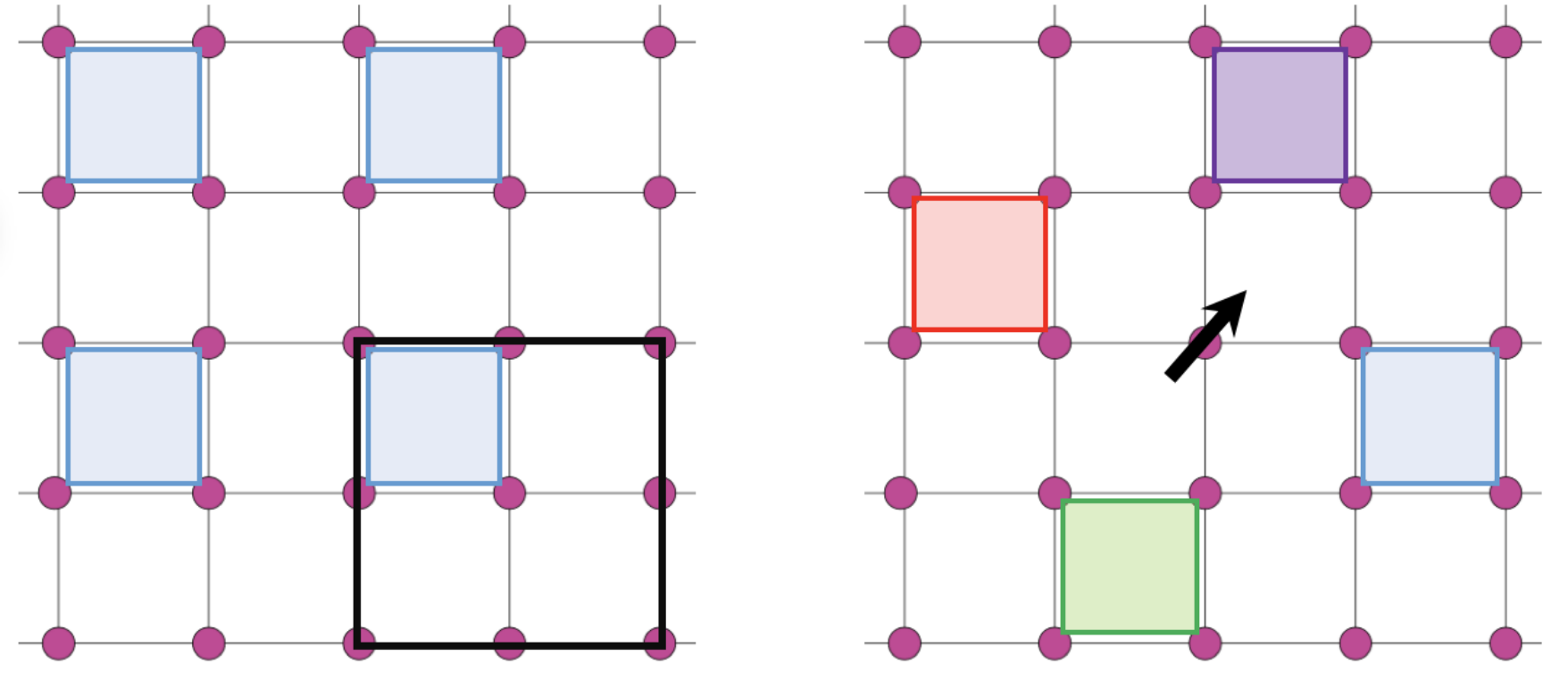}
  \caption{Left: VPS order which enlarges the unit cell by 4. Right: The vortex connecting four distinct VPS patterns carries a spinon.} 
\label{plaq1}
\end{figure}

In quantum magnets with geometry or quantum fluctuations, a zoology of paramagnetic states can emerge at low temperature. Beyond the well-know valence bond solid,
another widely observed paramagnetic crystalline phase is the VPS (valence plaquette solid) state which breaks $C_4$ symmetry and lattice translation $T_x,T_y$ for both directions. The VPS order enlarges the unit cell into four plaquettes, so there are four distinct VPS patterns related by site-centered $C_4$ rotation, as shown in Fig.~\ref{plaq1}.

\begin{figure}[h]
  \centering
      \includegraphics[width=0.55\textwidth]{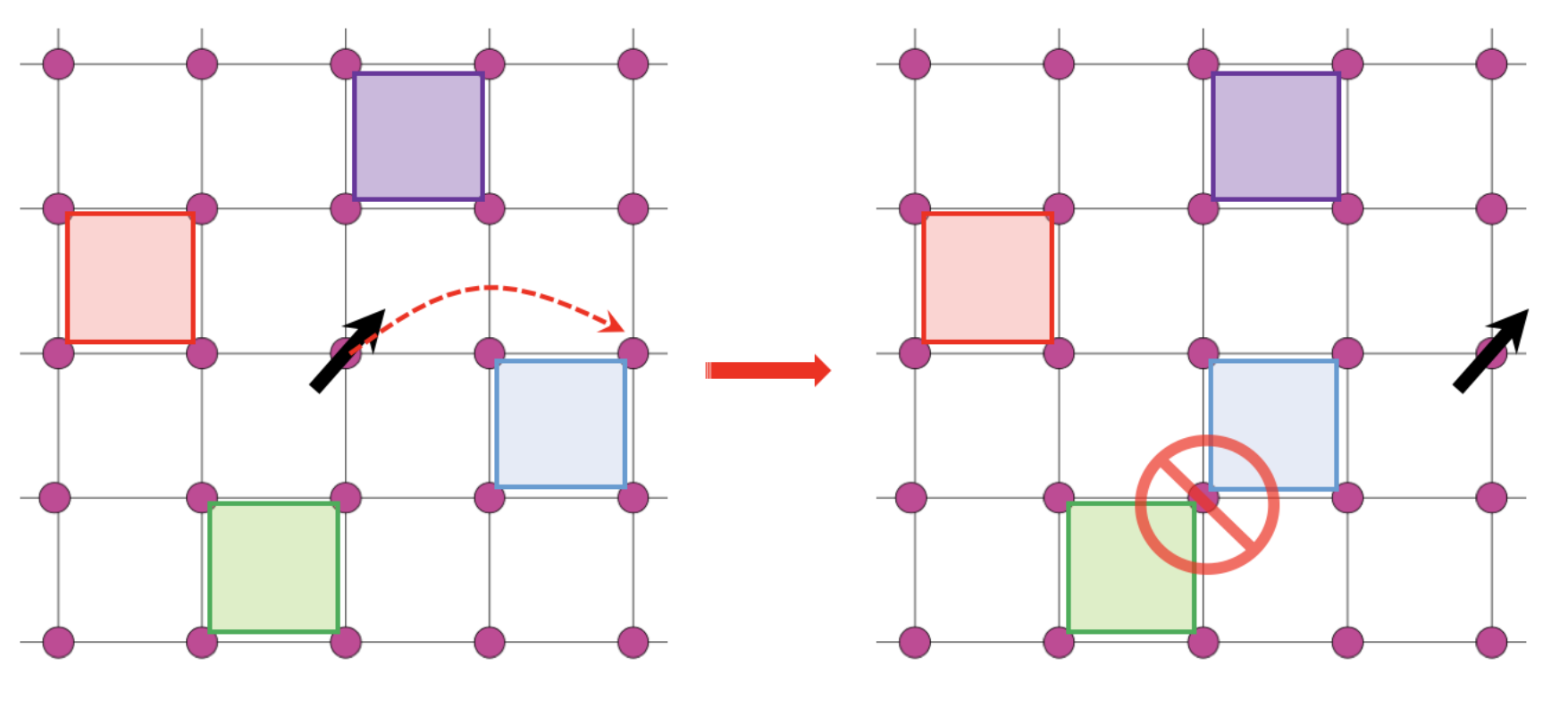}
  \caption{The spinon inside the VPS vortex has restricted mobility. It cannot move without breaking additional plaquettes.
} 
\label{plaq2}
\end{figure}

\begin{figure}[h]
  \centering
      \includegraphics[width=0.55\textwidth]{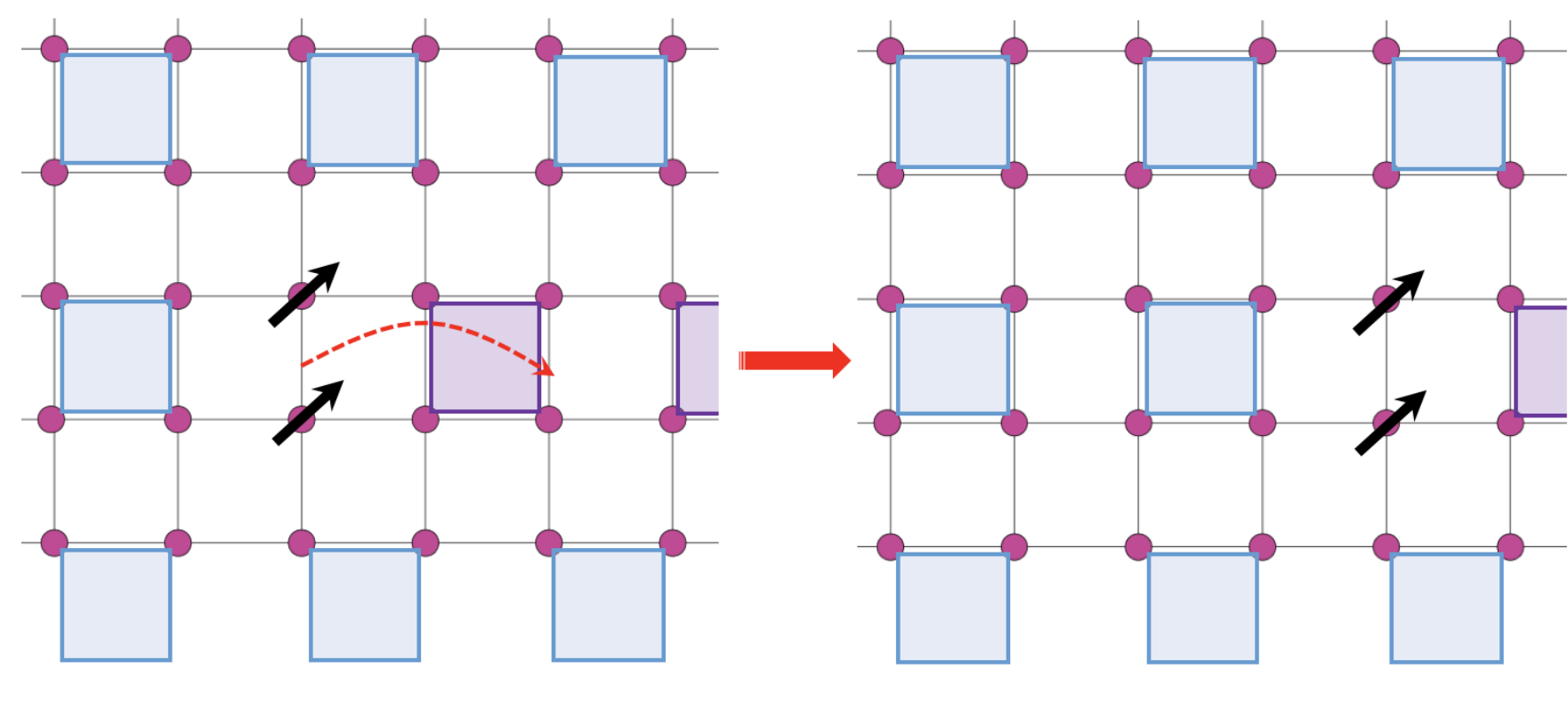}
  \caption{ A dipole can move along the stripe transverse to the dipole's orientation by exchanging position with a plaquette.
} 
\label{plaq3}
\end{figure}

In the plaquette crystalline phase, one can define the four distinct plaquette patterns as a $Z_4$ boson. During the quantum melting transition of VPS, the plaquette configuration tends to become disordered and the $Z_4$ vortex defect proliferates in the meantime.  The vortex defect of the VPS which intersects the four distinct plaquette configurations carries a free spinon as in Fig.~\ref{plaq2}.  As opposed to the VBS phase, where a spinon in the background of dimers can hop among sites by reconstructing the local valence bond configuration, a spinon in the background of plaquette order is frozen - it cannot move away from the original vortex center without breaking additional plaquettes, as depicted in Fig.~\ref{plaq2}.
In contrast, a pair of spinons living on the link between adjacent sites can hop along the stripe perpendicular to that link without breaking additional plaquettes, as depicted in Fig.~\ref{plaq3}. Such a spinon pair, which we refer to as a spinon dipole, is a $1d$ subdimensional particle which moves transversely to the dipole's orientation.  Based on these observations, the topological defect of the plaquette order displays restricted motion which exactly resembles the behavior of fractons. 

To make the connection between VPS defects and fractons precise, we introduce a higher rank gauge theory description for the valence plaquette order on a square lattice. The plaquette order can be mapped to a rank-2 symmetric tensor electric field defined at the center of each square as the following. 
\begin{align} 
E_{xy}(\bm{r})=(-1)^{i_{\bm{r}}} P(\bm{r})
\end{align}
where $P=1\,(0)$ corresponds to the valence plaquette occupancy (vacancy) on each square. The index $i_{\bm{r}}$ is the same as defined before. As opposed to the VBS state, where dimers can have two orientations corresponding to $E_x$ and $E_y$, the plaquette electric field is a single-component field, effectively a scalar. We can also define a conjugate variable $A_{xy}$, satisfying $[A_{xy}(\bm{r}),E_{xy}(\bm{r'})]=\frac{i}{2\pi}\delta_{\bm{r},\bm{r'}}$.  The operator $e^{\pm iA_{xy}}$ creates/annihilates a valence plaquette.  As each spin on the site is only entangled with one of the four adjacent plaquette clusters, one can define a Gauss's law for the rank-2 electric field as,
\begin{align} 
\partial_x \partial_y E_{xy}(\bm{r})=(-1)^{i_{\bm{r}}} (1-q(\bm{r}))
\label{gauss}
\end{align}
where $q(\bm{r})$ is the number of unpaired spinon at site $\bm{r}$.  As long as there is one plaquette adjacent to a site, there is no free spinon on that site.  If plaquettes are absent from all four squares surrounding the site, then there exists a free spinon charge at the center.  This Gauss's law is precisely the two-dimensional version of the Gauss's law seen in the fracton phase of matter described by a hollow rank-2 symmetric tensor gauge theory \cite{higgs1,higgs2}.  Due to the particular double derivative in Eq.~\ref{gauss}, the spinon number is conserved on each row and column of the system, so the theory respects an emergent subsystem $U(1)$ symmetry:
\begin{equation}
\int dx\,q = 1-(-1)^y\int dx\, (-1)^x\partial_x\partial_y E_{xy} = \textrm{const.}
\end{equation}
A similar equation holds in the $y$-direction. Due to the emergent subsystem symmetry, single spinon motion is prohibited. However, a pair of spinons, which we refer to as a dipole, can hop only along the stripe perpendicular to its orientation.

These mobility constraints, while they persist, can potentially inhibit the condensation of vortices and preclude a continuous transition from the VPS to the N\'eel antiferromagnet. Instead, the VPS melting transition can be driven by proliferation of spinon dipoles. In Ref.~\cite{plaquette}, it was demonstrated that a $2d$ VPS can melt into a stable gapless phase in the form of an algebraic bond liquid with algebraic correlations and long range entanglement.

\subsection{Hole-Doped Antiferromagnets}

\begin{figure}
    \centering
    \includegraphics[width=0.7\columnwidth]{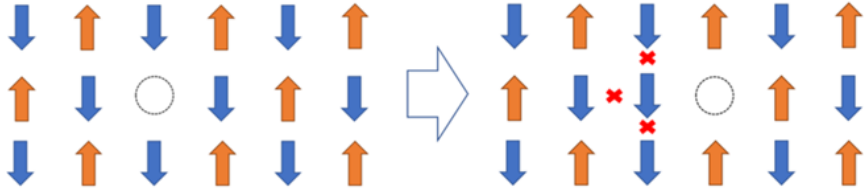}
    \caption{Motion of a hole through an Ising antiferromagnet is impeded by the creation of energetically costly spin misalignments.}
    \label{fig:hole}
\end{figure}

Another manifestation of fracton physics in a simple condensed matter system is in the familiar context of hole-doped antiferromagnets \cite{polaron}.  To see this, consider the motion of a hole through the background of an Ising antiferromagnet, as depicted in Figure \ref{fig:hole}.  As the hole moves via a sequence of nearest-neighbor hopping processes, it necessarily creates a series of misaligned spins, resulting in a large energetic barrier to motion.  This is in close analogy to the fact that a moving fracton must create energetically costly dipoles.  Furthermore, a bound state of two holes can move freely, without disturbing the antiferromagnetic background.  This provides an enticing hint that the physics of fractons is at play in the description of holes doped into an antiferromagnet.

This idea can be given firmer support by performing a perturbative analysis on the Hamiltonian for holes coupled to the antiferromagnetic background.  This type of system can be described by a form of boson affected hopping model, widely used in the theory of polarons.  In this type of model, hopping of one type of particle (which we typically assume to be fermionic) must be accompanied by the emission/absorption of another type of particle (which we take to be bosonic).  Schematically, such models take the form:
\begin{equation}
H = g\sum_{\langle ij\rangle}f^\dagger_i f_j (b_j^\dagger + b_i) - \mu_f \sum_i f^\dagger_i f_i - \mu_b \sum_i b^\dagger_i b_i
\end{equation}
where the first term represents the fact that hopping of a fermion requires either emission of a boson on the departure site or absorption of a boson on the arrival site.  (In the antiferromagnet, the $f$ particles would represent holes while the $b$ particles would represent magnon ($i.e.$ spin-flip) excitations.)  The fact that motion of a fermion is accompanied by creation/absorption of a boson is closely analogous to the fact that motion of a fracton is accompanied by the creation/absorption of an extra excitation (usually a dipole).  Indeed, by perturbatively integrating out the bosons through five orders in $g/\mu_b$, the authors of Ref. \cite{polaron} showed that the effective Hamiltonian for the fermions takes the schematic form:
\begin{equation}
H_f = -t\sum_i(f_{i+1}^\dagger f^\dagger_{i+2} + f^\dagger_{i-1}f^\dagger_{i})f_{i+1}f_i + V
\end{equation}
which features only a pair-hopping kinetic term, while $V$ schematically represents all potential/interaction terms.  To this order, we see that holes in a doped antiferromagnet are incapable of moving by themselves, but rather only move in pairs, in a manifestation of fracton physics.  A more precise analysis \cite{polaron} indicates that this model actually exhibits conservation of an appropriately defined dipole moment.  Furthermore, this conservation law leads to the usual phenomenology associated with fracton systems, such as slow thermalization and gravitational attraction.  In this case, the gravitation corresponds to the well-known magnon-mediated attraction between holes, which contributes to the formation of superconductivity \cite{dahm,gull}.  We therefore see that fracton phenomenology is on display in a very familiar setting, potentially with applications to the study of high-temperature superconductivity.

\subsection{Subsystem Symmetry Protected Topological Phases}

Symmetry plays a pivotal role in distinguishing phases of matter. The great majority of the exotic quantum phases lies in the interplay between symmetry and entanglement, which is known as 'symmetry protected topological phase'.  Then what are the possible quantum phases protected by subsystem symmetry? Do they exhibit similar protected gapless mode and symmetry anomaly on the boundary? In this section, we list several prominent examples of subsystem protected topological phase. 

The toy model we introduce here is a subsystem symmetric topological state\cite{you2018subsystem,devakul2018fractal,devakul2018classification,devakul2019strong,you2019higher,you2019highercr} with gapless edge modes, which we refer to as topological plaquette Ising model (TPIM).  The Hilbert space consists of Ising spins on sites of the square lattice.  For clarity, we will separate these into two spin flavours, $\sigma$ and $\tau$, located at the sites of the $A$ and $B$ sublattices, respectively.  The Hamiltonian is given by
\begin{align} 
H_\text{TPIM}=-\sum_{ijklm \in P_A} \sigma^z_i \sigma^z_j \sigma^z_k \sigma^z_l  \tau^x_m-\sum_{ijklm \in P_B} \tau^z_i \tau^z_j \tau^z_k \tau^z_l  \sigma^x_m
\label{topo1}
\end{align}
\begin{figure}[h]
  \centering
      \includegraphics[width=0.4\textwidth]{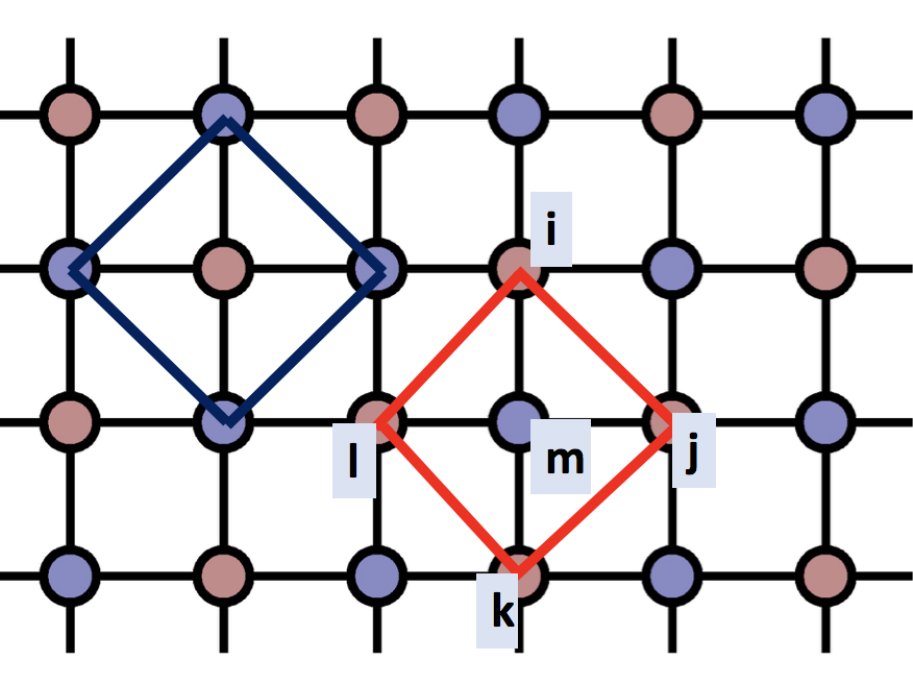}
  \caption{The terms in the TPIM Hamiltonian. The Pauli spins $\tau,\sigma$ live on the red/blue sites. The interaction $ \sigma^z_i \sigma^z_j \sigma^z_k \sigma^z_l  \tau^x$ involves the four $\sigma_z$ spins on the blue plaquette and the $\tau_x$ in the middle. The interaction $\tau^z_i \tau^z_j \tau^z_k \tau^z_l  \sigma^x$ involves the four $\tau_z$ spins on the red plaquette and the $\sigma_x$ in the middle. }
  \label{one}
\end{figure}
where the sum is over all $P_A$ ($P_B$), which refer to five-site clusters consisting of a site on the $A$ ($B$) sublattice and its four nearest neighbors, with each site labeled by $ijklm$ as illustrated in Fig~\ref{one}.  The first term is a sum over products of a $\tau^x$ and its four surrounding $\sigma^z$, and vice versa for the second.  As all local cluster-operators commute with each other, the Hamiltonian contains extensively many conserved quantities and is exactly solvable. Thus, the ground state of $H_\text{TPIM}$ can be described as a superposition of all possible $\{\sigma^z\}$ configurations, with the corners of each domain wall decorated with $\tau_x=-1$.  

In addition, the model has $Z^{sub}_2 $ symmetry, as the Hamiltonian commutes with the operators $\prod_{\text{diag}} \sigma^x$ and $\prod_{\text{diag}} \tau^x$ which flips $\sigma_z \rightarrow -\sigma_z$ or $\tau_z \rightarrow -\tau_z$ along a particular diagonal.

Next we will show that the SSPT paramagnet similarly has non-dispersing gapless boundary modes protected by the subsystem symmetry, which leads to a subextensive ground state degeneracy in the presence of an edge.



\begin{figure}[h]
  \centering
      \includegraphics[width=0.4\textwidth]{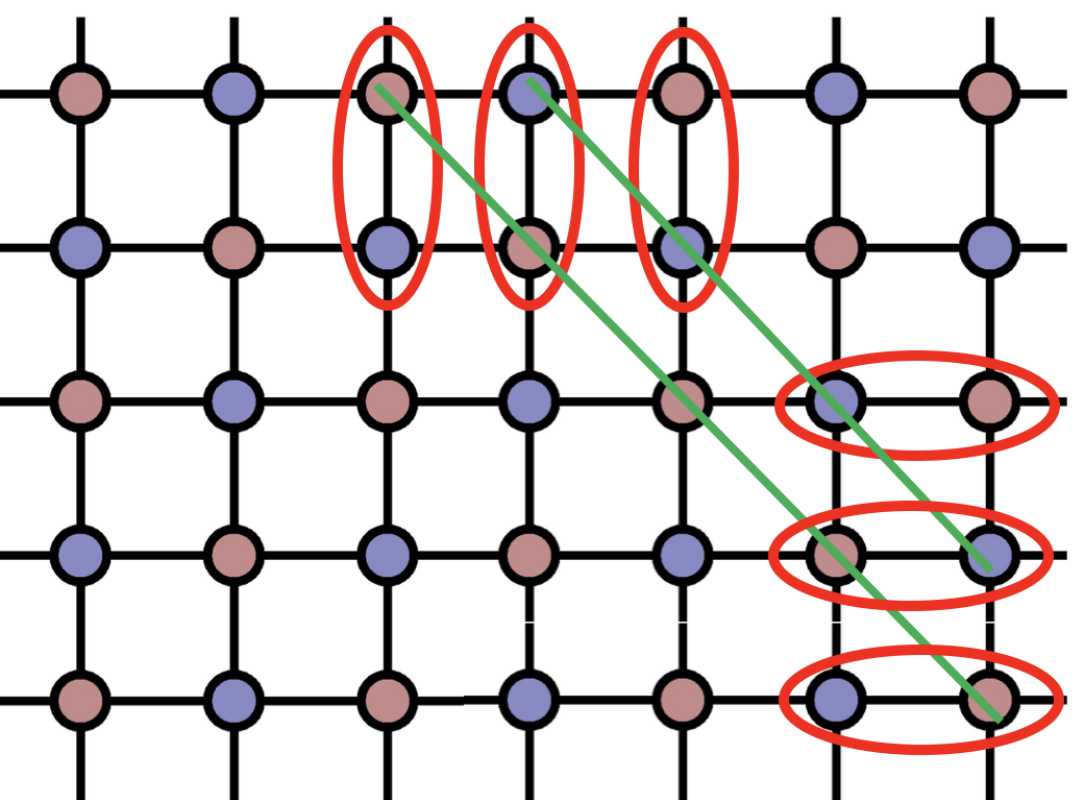}
  \caption{
  Red ovals show the physical spins that take part in the edge operators $\pi^\alpha_i$, and form a spin-$1/2$ degree of freedom at the edge.  
  The action of the subsystem symmetries (green lines) on the ground state manifold may be expressed in terms of such $\pi^\alpha_i$ operators.  
  Near a corner of the type shown here, the symmetry becomes a local symmetry, and the corresponding boundary modes can be gapped out.}
  \label{three}
\end{figure}

Consider an edge as shown in Fig.~[\ref{three}]. We can pick two-spin clusters (red ovals in Fig~\ref{three}), which create an effective spin 1/2 degree of freedom on each site along the edge. 
To see this, we define the three edge operators for each cluster on even sites,
\begin{align} 
\pi^x=\sigma^z \tau^x,\pi^y=\sigma^z \tau^y, \pi^z= \tau^z
\label{edge1}
\end{align}
and likewise, for odd edge sites with $\sigma$ spin at the surface, we have
\begin{align} 
\pi^x=\tau^z \sigma^x,\pi^y=\tau^z \sigma^y, \pi^z= \sigma^z
\label{edge2}
\end{align}
These operators satisfy the Pauli algebra on the surface, and commute with the bulk Hamiltonian $H_\text{TPIM}$.
By counting degrees of freedom, we can see that there exists a $2^L$-fold degenerate ground state manifold arising due to the presence of the edge of length $L$.

This edge degeneracy in fact cannot be broken with local interaction while preserving all subsystem symmetries, and leads to a completely flat-band dispersion along the edge.

\section{Conclusions and Outlook}
 \label{sec:conc}

In this review, we have given a bird's eye perspective on the field of fractons, which is an exciting new frontier for condensed matter physics.  Fractons not only represent a fundamentally new type of emergent quasiparticle with striking properties, but also draw connections between a variety of seemingly unrelated topics, from gravity and elasticity to higher-order topological insulators and hole-doped antiferromagnets.  While we have covered a wide range of topics, there have been many other exciting advances in the field which we have not discussed here, and we refer the interested reader to the literature for more information \cite{he2019lieb,you2019higher,devakul2019classifying,muhlhauser2019quantum,circuit,tantiv,you2019non,bulmashgauging,spectra,madeconfined,parton,tian2018haah,juven3}.

Beyond established results, however, there are also numerous open questions in the field of fractons, which has entered into a new stage of maturity.  These open questions range from the practical to the highly abstract.  As always, one important line of research is the search for more experimentally-relevant spin models which may be realized in actual materials exhibiting frustrated magnetism.  It will also be important to develop more experimental signatures of fractons in spin systems, particularly for gapped models.  However, recent developments have made it clear that fracton physics is a much broader paradigm than its humble beginnings in exactly solvable spin models.  Fractons are already known to be realized in a diverse set of systems, such as elasticity theory, plaquette paramagnets, hole-doped antiferromagnets, and more.  As such, it is natural to ask what other platforms may host fractons, and how fracton physics is concretely manifested in experimental signatures.

Given that fractons are on the cusp of physical realization, it is also important to ask what we will do with fractons once we have them.  How can we practically manipulate fractons in some useful way?  It has been widely suggested that the properties of fractons will be useful for the purpose of quantum information storage \cite{haah,bravyi,terhal,parallel}, but we lack any concrete roadmap for the precise implementation of this proposal.  Much more work will be required to figure out how to usefully store and manipulate quantum information using a fracton system.  It is also unclear whether or not the mobility restrictions of fractons can be harnessed for constructing any other sort of useful quantum devices.

On the more abstract side of things, the number of open questions is still remarkably large.  One important line of research is a push towards a complete classification of fracton systems, with a full characterization of all statistical processes.  There has also been only limited exploration of fractons in fermion systems, and the known models all have natural analogues in boson systems.  Are there examples of intrinsically fermionic fracton models?  For example, can fermion systems give rise to tensor gauge theories with half-integer higher-spin gauge modes?  Another interesting question is what we can learn about real gravitational systems from the connection between fractons and gravity.  Can fracton physics provide new insights into more traditional gravitational theories?  Can fracton models be used to simulate more complicated gravitational phenomena, such as black holes?

This list of open questions is far from exhaustive, and new topics in fracton physics are being discovered at a rapid pace.  It seems fair to say that the field of fractons still has many surprises in store over the coming years.  We hope that this review will serve as a useful introduction for the next generation of fracton researchers, who will surely take the field in many exciting new directions.

\section*{Acknowledgments}
We are grateful for helpful discussions with Kevin Slagle, Wilbur Shirley, and Shriya Pai. X.C. is supported by the National Science Foundation under award number DMR-1654340, the Institute for Quantum Information and Matter at Caltech, the Walter Burke Institute for Theoretical Physics at Caltech and the Simons Foundation through the collaboration on Ultra-Quantum Matter. Y.Y. is supported by PCTS Fellowship at Princeton University. This material is based in part (M.P.) upon work supported by Air Force Office of Sponsored Research under grant no. FA9550-17-1-0183. YY, MP are supported in part by the National Science Foundation under Grant No.NSF PHY-1748958(KITP) during the Topological Quantum Matter program. This work was partly initiated at Aspen Center for Physics, which is supported by National Science Foundation grant PHY-1607611. 

%

\end{document}